**Unstructured Grid Dynamical Modeling of Planetary Atmospheres using planetMPAS: The Influence of the Rigid Lid, Computational Efficiency, and Examples of Martian and Jovian Application.**

Yuan Lian[a] and Mark I. Richardson[a]

[a] *Aeolis Research, Chandler, AZ 85224, USA*

*Corresponding author*: Yuan Lian, lian@aeolisresearch.com

**Abstract**

We present a new planetary global circulation model, planetMPAS, based on the state-of-the-art NCAR MPAS General Circulation Model. Taking advantage of the cross compatibility between WRF and MPAS, planetMPAS includes most of the planetWRF physics parameterization schemes for terrestrial planets such as Mars and Titan. PlanetMPAS also includes a set of physics that represents radiative transfer, dry convection, moist convection and its associated microphysics for the Jovian atmosphere. We demonstrate that, despite the rigid-lid approximation, planetMPAS is suitable to simulate the climate systems in Martian and Jovian atmospheres with potential application to slow-rotating planets such as Titan. Simulations using planetMPAS show that the new model can reproduce many aspects of the observed features on Mars and Jupiter, such as the seasonal CO2 cycle, polar argon enrichment, zonal mean temperature, and qualitative dust opacity on Mars, as well as the equatorial superrotation and banded zonal wind patterns on Jupiter.

**1. Introduction**

Dynamical modeling of planetary atmospheres and climates is now largely undertaken with general circulation (or global climate) models (GCMs) that explicitly simulate synoptic scale motions. (The definition of the "synoptic" depends upon the size and rotation rate of the object, but for Earth and Mars we can usefully define it as motions with lateral scales >500 km, and we will use "global scale" and "synoptic scale" as interchangeable terms). Such models have been used to examine the atmospheres of Venus (*e.g.*, Lee et al. [2007]; Lee and Richardson [2012]; Lebonnois et al. [2016]; Sugimoto et al. [2019]), Mars (*e.g.*, Leovy et al. [1969]; Pollack et al. [1981]; Haberle et al. [1993]; Hourdin et al. [1993]; Wilson and Hamilton [1996]; Forget et al. [1999]; Lewis et al. [1999]; Richardson et al. [2007]; González-Galindo et al. [2013]; Shaposhikov et al. [2019]), gas giant planets (Lian and Showman [2008; 2010]; Schneider and Liu [2009]; Palotai and Dowling [2008]), Titan (Mitchell et al. [2011], Schneider et al. [2012], Newman et al. [2011]), Pluto (Toigo et al. [2015], Forget et al. [2017]), and extra-solar planets (*e.g.,* Showman et al. [2009]; Wolf et al., [2017]). A more limited body of research has seen the application of numerical dynamical models operating on smaller meso- and microscales (for Earth and Mars we can define these as 10-500 km and <10 km, respectively). Meso- and microscale models employ similar dynamical equations to those of GCM's, but often with fewer approximations such that non-hydrostatic and compressible motions are also admitted. Most application of meso/microscale models has been to Mars (Tyler et al. [2002]; Rafkin et al. [2001]; Toigo and Richardson [2002]; Richardson et al. [2007]; Spiga and Lewis [2010]), though they have also been applied to Venus (Lefevre et al., [2020]) and Titan (Barth and Rafkin [2010]). Almost all of these models are planetary applications of existing dynamical models built for simulation of the Earth's atmosphere. As such, most of these studies are only made possible by the investment in the development of complex numerical models by terrestrial research laboratories such as the Geophysical Fluid Dynamics Laboratory (GFDL), the NASA Goddard Institute for Space Sciences (GISS), Program in Atmospheres , Ocean and Climate at Massachusetts Institute of Technology (MIT PAOC), the Laboratoire Meteorologie Dynamique (LMD), the Met Office, and the National Center for Atmospheric Research (NCAR) for application to research problems that range from terrestrial climate change, to weather and renewable energy resource forecasting.

Most current global planetary atmospheric models make use of spatial (as opposed to spectral) horizontal discretization on structured numerical meshes, most commonly a rectangular mesh on a Plate Carree simple cylindrical map projection (also known as polar-spherical grid or traditional longitude-latitude grid). With these meshes, the meridional grid lines converge at poles yielding the well-known "polar convergence problem" in which the zonal extent of grid boxes become progressively smaller as the pole is approached. To maintain numerical stability, these polar points either require a reduction in the model timestep or a filtering process to remove fast motions, such as polar Fourier filtering. The polar convergence is thus a problem both for reasons of computational efficiency and because of the "damage" the filtering schemes can do to the numerical solutions. Some improvement upon this problem has been made with cubed-sphere models that have less, but still notable, deformation where the square faces of the cubed-sphere grids meet, *i.e.*, cube-sphere corners (Adcroft et al. [2004]; Putman and Lin [2007]). The polar filtering in the simple cylindrical grids can be problematic for tracer advection (Lian et al. [2012]) while the grid distortion at the cube-sphere corners can result in strong numerical oscillations "locked" to these locations under certain dynamic conditions, such as occur on slowly rotating objects like Titan. The simple-cylindrical grids are also non-optimal in terms of computational efficiency since the distorted nature of the grids means that more points are needed to cover the entire spherical surface for a given resolution than would be required if the grid points were evenly spaced (*i.e.*, than if they were geometrically equidistant). Likewise, the polar convergence of cylindrical grids often requires smaller model time step (lower computational efficiency) to avoid the violation of Courant-Friedrichs-Lewy (CFL) criterion. These grids are also not optimal from the perspective of parallel computation due to the need to gather full zones of variables for Fourier filtering.

Structured-mesh mesoscale models generally do not exhibit as extreme a distortion of the numerical grid as global "longitude-latitude" models, nor do they require spatial filtering, but they generally require the nesting of numerical meshes "one-within-the-other" and usually require a separate model to drive the boundary forcing at the largest scale. The wall boundaries are problematic due to the sharp boundary in resolution (and sometimes in numerics at the parent/daughter domain interfaces) and due to the fact that in some cases the nesting is one-way (large scales influence the small, but there is no feedback from the small to the large). Even when the numerics are the same at all scales and the feedback is two-way, the sharp resolution jump can be problematic in three ways:

   1. Fluid dynamical systems that are advected into the high-resolution domain from the parent by definition lack the high-resolution fluid structure that would characterize a real system (*e.g.*, turbulent boundary layer structures that should be present within the inflow for nested LES), and hence the simulated flow in the higher-resolution domain may possess the wrong high-frequency dynamical structure compared to reality (*e.g.*, Moeng *et al*. [2007]).

2. Flows and waves that were destined to travel out of a high-resolution domain into a lower resolution parent can be partially reflected back from the nest wall due to abrupt change of spatial resolution.

3. The infrastructure needed to deal with the nesting process is generally computationally expensive, *i.e.*, the interpolation processes between a fine nest and its parent domain.

Problems with structured grids have been recognized for a long time but have remained in use for a few reasons. First, the design of numerics for unstructured grids is a non-trivial task, and second, computational power had for a long time limited model application to relatively low resolution simulations where there is a relatively small performance advantage of more exotic unstructured grids compared to more traditional structured grid models. However, the parallel computing capability has increased massively over the past 15-20 years and the desire within the terrestrial community to advance towards global mesoscale modeling (also referred to as so-called global "convection-permitting modeling") has increased interest in more efficient unstructured grid dynamics. The interest in convection-permitting modeling for the Earth is driven primarily by the importance of cloud and moist convective processes for accurate prediction of weather and for the simulation of the global climate system (*e.g.*, Judt [2020]). Of particular interest for this paper has been the development of the atmospheric version of the Model for Prediction Across Scales (MPAS), primarily at NCAR (Skamarock et al. [2012]). As described in greater detail, below, MPAS (mpas-dev.github.io) implements an arbitrary unstructured mesh of grid points using centroidal Voronoi tessellation (CVT), also referred to as spherical CVT (SCVT) when projected into a sphere, or simply as an unstructured Voronoi mesh. The grid can be designed to provide arbitrary coverage of the planet, from meshes that cover the planet with nearly uniform spacing, to providing gradual zooming on specific locations of interest. MPAS can also be run with a limited area domain using either forced or idealized boundary conditions (Skamarock *et al*. [2018]). Most significantly, the numerics have been designed to provide very high-fidelity tracer conservation and to scale very efficiently across shared and distributed memory on massive parallel supercomputers (Heinzeller *et al*. [2016]).

Just as dynamical processes and forcing across the mesoscale/synoptic scale boundary is the frontier in modern weather and climate study for the Earth, many of the most significant problems in planetary atmospheres are also faced with similar cross-scale challenges. For example, dust storm dynamics on Mars involve storm growth from the smallest mesoscales (or even microscales) to global phenomena, via mechanisms that are both poorly understood and very challenging to numerically examine without the ability to simulate dynamical interactions across the scales. For Titan and the gas giant planets, moist convection poses analogous problems with very similar physical basis to that of the moist convection problem for the Earth. The availability of the MPAS atmospheric model thus has great potential for studying nonhydrostatic and fully compressible planetary atmospheres across the scale spectrum from meso- to global scales.

The purpose of this paper is to provide an initial assessment of the suitability of the MPAS model for planetary applications (planetMPAS). We specifically convert MPAS following the planetary framework of the planetWRF system (Richardson *et al.* [2007]), which was in turn developed from the NCAR Weather Research and Forecasting (WRF) model (Powers *et al.* [2017]). One feature of the MPAS atmospheric model is its use of a height-based terrain following vertical coordinate, which has not typically been used for planetary models. For Mars, where both the bulk mass and the global mean temperature of the atmosphere change significantly over the annual cycle, this choice may not be ideal. There are other potential planetary applications in which the domain average temperature may change significantly over the course of a simulation. We address the kinds of planetary environments to which the height-based model can be applied. For comparison with existing global planetary atmospheric modeling, we also examine the computational efficiency and scaling of a Martian version of MPAS (MarsMPAS) on the NASA HECC Pleiades supercomputer and compare this with the latitude-longitude GCM MarsWRF on the same processors and using the same model physics. We finally provide some initial examples of MPAS applied to Mars and Jupiter (JupiterMPAS) to demonstrate the ability of MPAS to yield simulations of at least comparable fidelity to existing global model simulations at comparable resolutions. We will specifically show that as regards tracer transport in the Martian polar atmosphere, MarsMPAS is significantly more capable than existing GCMs.

In Section 2, we provide an overview of MPAS with a specific focus on features useful for planetary application and a brief discussion of the steps needed to modify MPAS for planetary implementation. Significantly, the NCAR MPAS atmosphere model was designed to use 'physics' routines nearly unmodified from the WRF model framework. Since WRF has already been converted to planetary use and given that a large library of physics routines for a variety of planetary applications already exists within the planetWRF system, conversion of MPAS for planetary use (to create the planetMPAS model) has been much more straightforward than the conversion of prior models. In Section 3, we provide a discussion of the "pressure cooker effect" which has been raised as a concern for height-based vertical grids (Rafkin, personal communication, 2016), such as MPAS, that necessarily possess a rigid upper lid. We provide a detailed assessment of the problem from an idealized perspective and then show MarsMPAS output.

In Section 4, we examine the computational performance of the MarsMPAS model on modern parallel computers and compare it with the performance of the MarsWRF and Mars MITgcm models. In Section 5, we described the conversion of MPAS to Mars, including an initial assessment of the zonal mean thermal structure, seasonal $CO_2$ cycle and zonal mean dust distribution developed by the model. We also describe significantly improved simulation of the southern polar winter argon abundances compared to prior GCM modeling (Lian et al. [2012]). In Section 6, we describe the conversion of MPAS to Jupiter, including an initial assessment of the equatorial superrotation and belt-and-zone banded jet structure. Finally, we provide a summary in Section 7.

## 2. Model Description

MPAS provides a fully compressible and nonhydrostatic numerical simulation of a three-dimensional atmosphere (Skamarock *et al.* [2012]). In this work we have used MPAS v7.0. The model system can be divided into four broad categories of code:

1) The dynamical core (or the "dynamics") that simulates fluid motions by solving Navier-Stokes equations under spherical geometry in a rotating frame.
2) The "physics" suite that represent non-dynamical or processes unresolved on the model mesh, such as radiative heating and boundary layer convection (in this paper we do not consider the feasibility of performing Large Eddy Simulation in MPAS, though MPAS can be run in this way with high-resolution mesh in a limited-area domain).

3) The code framework or infrastructure necessary to drive both the dynamics and the physics, to exchange information between the dynamics and the physics, and to handle initialization, file input/output, and code parallelization for use on modern multiprocessor computers.

4) Separate offline codes needed to configure existing MPAS meshes provided by NCAR and generate the initial model prognostic fields and boundary conditions (for limited-area simulations). Note that generating MPAS SCVT meshes can be complicated and computationally expensive.

Each component of the system is modified to some degree for planetary applications. In the remaining part of this section, we provide a very brief orientation to the MPAS system and outline the modifications that were necessary to yield planetMPAS. The description of the model is not meant to repeat or replace the detailed and authoritative description by Skamarock *et al*. [2012]. Instead, we here provide only enough information to a high-level overview for planetary scientists and sufficient details to understand the necessary planetary changes.

## 2.1 The MPAS Dynamical Core

MPAS atmosphere dynamical core solves a system of fully compressible, non-hydrostatic equations in a modified height-based coordinate ($\xi$-coordinate) (Klemp [2011]) that closely resemble the pressure-based terrain-following sigma or sigma-hybrid coordinates used in WRF (Skamarock *et al.* [2012]; Powers *et al.* [2017]). The prognostic equations, such as for vertical momentum, heat and mass transport, are written in flux form to achieve better conservation properties. The horizontal momentum equation is formulated in vector-invariant form that suits various types of structured or unstructured grids where cell edges are not aligned with east-west and north-south directions. For instance, the horizontal momentum equation in MITgcm cube-sphere grid is also formulated in the vector-invariant form to ensure that the vectors remain unchanged during coordinate transformation (Adcroft *el al.* [2004]). Due to the changed vertical coordinate compared to WRF, density and potential temperature are prognostic, with pressure being diagnostic. The equations integrated by MPAS are described in details by Skamarock *et al.* [2012].

The governing equations are solved using the finite volume method (FVM), a conservative method that applies to complex grid geometry. FVM has superior tracer conservation properties compared to the finite difference method (FDM) used, for example, by WRF. This is particularly important when studying tracer transport problems, which can be important in planetary atmospheres (*e.g.*, the enhancement of Argon in Mars' atmosphere (Lian *et al.* [2012])). The time integration in MPAS adopts third-order Runge-Kutta method with explicit time splitting that integrates gravity waves and acoustic waves on time steps smaller than main model time step (Klemp *et al.* [2007]; Wicker and Skamarock [2002]). The time-splitting method, while slightly increasing the computational cost, can effectively reduce numerical instabilities caused by fast propagating waves that violate the CFL criterion when using a large model time step. The differential equations are spatially discretized on a staggered C-grid CVT mesh. As shown in Figure 1, on this mesh the horizontal momentum is placed on the cell edges and scalars such as temperature, density, tracers and kinetic energy are placed at cell centers (Thuburn *et al.* [2009]; Ringler *et al.* [2010]). The distribution of these cells to cover a sphere, which is an SCVT global mesh, is shown in a very low-resolution example (roughly 250km grid spacing or nearly 4 degrees at the equator) for Mars in Figure 2. The low-resolution mesh is shown so that the arrangement of the individual polygonal cells can be discerned on the global projection.

The MPAS atmospheric model implements several numerical viscosities and wave damping mechanisms to suppress numerical instabilities. Two-dimensional Smagorinsky viscosities (Smagorinsky [1963]) or fixed horizontal eddy viscosities are used to horizontally diffuse grid-scale noise in the temperature and momentum fields. By default, a second-order Smagorinsky viscosity is active for all simulations and a fourth-order hyper-viscosity is tunable by users. Alternatively, fixed second-order eddy viscosities/diffusivities can be applied to mix temperature and momentum in both the horizontal and vertical directions. Because MPAS uses a height-based coordinate to formulate the non-hydrostatic motions, the top boundary becomes a rigid lid where upward-propagating gravity and acoustic waves are reflected. An implicit, user-configurable Rayleigh damping of the vertical momentum equation is constructed to damp the amplitudes of upward-propagating waves while retaining the wave structures (Klemp *et al.* [2008]). For Mars, this is particularly important over regions where waves are generated by sharp terrain, such as Olympus Mons, by dust devils or by $CO_2$ moist convection. Additionally, divergence damping is used in the momentum equations to filter out acoustic wave modes (Klemp *et al.* [2018]; Skamarock and Klemp [1992]). The divergence damping combined with an off-center implicit vertical transport scheme can enhance the numerical stability when fast traveling waves are present, but it can also suppress the magnitude of high-order atmospheric thermal tides in the Martian atmosphere (Lian et al. [2022]).

The advection schemes for scalars in MPAS are similar, but fewer in number, to those available in WRF. These conservative schemes are based on second and third order Runge-Kutta time (RK2 or RK3) integrations and utilize both centered and upwind-biased spatial discretization for scalar advection terms expressed in flux form. Wicker and Skamarock [2002] showed that, for compressible atmospheric flows, these advection schemes are numerically stable with a large time step (even when the Courant number is greater than 1) and are more accurate than the schemes based on the leapfrog time integration. In the current MPAS version, the advection schemes are limited to third-order upwind spatial discretization with the RK3 time integration.

Most numerical tracer advection schemes will inadvertently produce unphysical negative tracer mixing ratios and even negative atmospheric densities. To address this issue, the MPAS atmospheric model implements a monotonic flux limiter, which adjusts the inflow and outflow mass fluxes of a control volume to ensure that the tracer mixing ratio or density in this control volume will be bounded by the initial minimum and maximum inflow values (Thuburn [1996]). In other words, the monotonic flux limiter is positive-definite, meaning that tracer mixing ratio or density remains positive if it was initially so (Skamarock [2006]; Skamarock and Weisman [2009]; Durran [1999]; Thuburn [1996]). The monotonic flux limiter is applied to the RK3 based advection schemes.

The MPAS dynamical core is mostly unmodified for planetMPAS. The only modifications that have been implemented at the time of writing have been: 1) source/sink terms for atmospheric density that allow bulk $CO_2$ condensation / sublimation for the case of MarsMPAS, 2) a secondary tracer that contributes to atmospheric mass besides the default active tracer water vapor (*e.g.*, argon for Mars, ammonia for Jupiter and Saturn), 3) an optional variable gravity that is a function of radial distance from the center of planet, and 4) a vertical component of the Smagorinsky viscosity (Smagorinsky [1993]). Constants in the MPAS dynamical core are cleanly specified in a single location in an included module (part of the model common framework). We have followed convention used in WRF, MPAS, and planetWRF of using c-preprocessor directives to allow specific model versions to be created at compile time and generate the appropriate planetary constants sets for planetMPAS. Some of the dynamical core filtering / diffusion settings have been modified for different planetary applications, as described in the planet-specific setups described in Sections 5 and 6.

## 2.2 MPAS Physics

MPAS has been designed from the outset to use the same physics routines as the WRF model, and both models have been developed by the same group at NCAR. This means that planetary physics modules developed within the planetWRF framework work with little or no modification within MPAS. Both WRF and MPAS integrate these physics routines into the model using intermediate-level physics drivers. These driver routines, which are different between MPAS and WRF, provide an interface between the parent model (dynamical core) and the physics routines such that the physics modules do not need to "know" anything about the parent model grid structure, grid arrangement, model input/output handling, or parallelization. In essence, the WRF/MPAS physics routines provide vertical column models of physics processes, passing tendencies back to the parent model. Thus, only the model-specific driver layer code needs to be customized for MPAS versus WRF. The Common Community Physics Package (CCPP), currently in development for a wide range of models including WRF and MPAS (Heinzeller *et al*. [2020]), takes this commonality one step further by formalizing the physics and driver routines as an integrated, stand-alone package with a common interface that will allow even more ready interchangeability between WRF and MPAS. The commonality of the physics, even for the current state of the models without the CCPP, means that WRF and MPAS can be thought of as alternate dynamical cores available for use with a common pool of physics. This approach, combined with the existence of the planetWRF physics code (git / github) repository, has made conversion of MPAS for planetary usage much easier than for any prior atmospheric general circulation model.

## 2.3 MPAS Framework

Only very minor modifications of the MPAS framework are required. Model constants in MPAS are retained in centralized modules where they are defined for use everywhere in the code for both dynamics (*e.g.*, planet radius, gravity, rotation period and major gas constants) and physics (*e.g.*, trace gas properties such as latent heat and other constants related to microphysics/radiation). For planetMPAS, we follow the WRF/MPAS convention of using pre-processor directives to allow all the planetary code (*i.e.*, code for all the different planetary instances) to reside in a single code base and to be selectable at compile-time. Clocks-and-calendar issues are handled by integrating the model using SI time, and then calculating planet-specific clocks-and-calendar values (like orbital position and strength of solar forcing) inside of the radiation driver code. Planetary rotation rates and other clocks-and-calendar constants are managed in the same way as the other constants in the central constants module.

**2.4 MPAS Initialization**

PlanetMPAS is initialized from startup files generated by codes originally designed for idealized and real-atmosphere test cases for the MPAS system. The initialization code allows surface data to be averaged or interpolated onto the MPAS mesh, employing generalized routines common with planetWRF (Toigo *et al.* [2012]). This surface data includes topographic properties (*i.e.*, terrain elevation, terrain slopes, orographic asymmetry and variance), albedo, emissivity, roughness and thermal inertia. For example, Mars terrain from the Mars Orbiter Laser Altimeter (MOLA) is shown on a low-resolution mesh in Figure 2. The initial state takes a user-specified surface pressure (hence the total column air mass) and thermal structure (temperature-height profile or temperature-pressure profile) and assumes the atmosphere initially to be at rest (stationary atmosphere) and uses this information to generate a hydrostatically balanced density structure. Both passive tracers (such as Argon and Nitrogen) and active tracers (such as water) can also be initialized via the initialization routines. For the Mars case, the passive tracer Argon is currently initialized to a mass mixing ratio of 0.0145 kg/kg that is uniformly distributed throughout the atmosphere (Lian *et al.* [2012]; Sprague *et al.* [2007]). By varying the initial surface pressure, the total mass of $CO_2$ and Argon is adjusted to about $2.85 \times 10^{16} kg$ to reproduce the Viking Lander pressure cycles properly (Guo *et al*. [2009]). Additionally, results from WRF and other GCMs can also be used to generate initial and lateral boundary conditions for limited-area simulations in MPAS via horizontal and vertical interpolations of WRF Preprocessing System (WPS) data (in longitude-latitude format) to MPAS meshes (Skamarock *et al.* [2018]).

**3. Rigid Lid Considerations**

Prior global planetary atmospheric dynamical models have typically used pressure-related vertical coordinates. Most commonly, the near surface atmosphere has been treated using terrain-following coordinates that are closely related to the sigma coordinate, $\sigma = p/p_s$, where $p$ is the pressure at the level center and $p_s$ is the local surface pressure. At altitude, the terrain following coordinate is either maintained or transitioned to a true pressure coordinate. Such pressure-related vertical coordinates do not allow transport of mass out of the vertical domain, but they do allow the geopotential height of the model top to deform upwards or downwards in accord with the column integrated temperature. As such there is a free upper boundary on the modelled atmosphere. The surface pressure is prognostic in such models, and the interior fluid pressure is either diagnostically reconstructed from the vertical grid structure and the surface pressure (for hydrostatic models) or from a combination of this reconstruction along with prognostic calculation of the nonhydrostatic perturbation pressure structure.

MPAS, akin with another atmospheric model widely applied to planetary atmospheres, the Regional Atmospheric Modeling System (RAMS) (Rafkin and Micheals, 2019), instead uses a height based vertical coordinate. In such a model, density is prognostic and the surface and fluid interior pressures are diagnosed from the density and temperature using the equation of state. With the height of the ground and that of the model top both fixed, such height-based models do not possess a free upper boundary, unlike the pressure-based models. One concern that has been raised in planetary applications of such height-based vertical coordinate models is how the diagnosed surface pressure behaves within a fluid that is essentially "trapped" between two fixed rigid layers (the surface and the rigid model lid) when the bulk atmospheric temperature or mass changes significantly [Rafkin, personal communication, 2016]. This is not a significant issue for the Earth, where the mean global temperature remains within a few K over the seasonal cycle and the bulk atmospheric mass does not vary significantly. However, the bulk lower and middle atmospheric temperature on Mars varies by roughly 20 K between late northern spring, when Mars is near aphelion, and the southern summer solstice, when Mars is nearer perihelion; and the air temperature cycle can be another 10 K or more larger in years with global dust storms (*e.g.*, Liu *et al*. [2003]). The bulk atmospheric mass also varies by about 25% over the annual cycle due to exchange with the seasonal $CO_2$ ice caps.

The nature of the potential problem can most readily be visualized by considering a pot of air. If a lid is placed on the pot and sealed, the initial temperature and pressure within the pot are the same as those of the near surface atmosphere. If the atmosphere outside the pot is heated (laterally uniformly), the air can rise such that the density can come into balance with the temperature with a new, more distended hydrostatic vertical profile of pressure. Since no mass is added to the air column, the surface pressure does not change as the air column is heated (see Appendix A of Richardson and Newman [2018] for a more detailed discussion). For the sealed pot, however, an increase of temperature cannot be accommodated with a decrease of density since the volume and mass of air within the pot are fixed. The density is thus required to remain constant. Instead, an increase of temperature in the seal pot necessarily causes the air pressure to increase (following the ideal gas law). The higher the temperature is increased the higher the pressure becomes, even though the initial pressure was determined by the atmospheric surface pressure. The "pot" thus becomes a "pressure cooker". The deviation of the pot surface pressure from the free column surface pressure is a result of the atmosphere within the pot not being allowed to expand vertically as it is warmed. The magnitude of the effect is sensitive to how deep the pot is. For a very shallow pot, where the pot depth, $z \ll H$, where H is the scale height, the effect is largest as all no expansion of the column is allowed at all. If the pot is very tall, such that the top pressure $P_T$ is almost 0 ($z \gg H$), then the sealed pot acts effectively like the free atmosphere. In intermediate cases, mass can redistribute vertically, but is constrained by the constant column-average density within the domain height, z.

The pot becomes more applicable to modeling of a real atmosphere if we consider the pot to be laterally extended and, in the limit, where the notional "pot" had no sides, but instead it forms a pair or concentric spherical shells around the planet, and where the domain height z is O(H). In this case, the configuration of the "pot" is the same as a shallow rigid lid model. Hence, the foregoing concern that the surface pressure would incorrectly vary with global average thermal state in a rigid lid model is "the pressure cooker effect" (as coined by Rafkin, personal communication, 2016).

The "pressure cooker effect" for a vertically resolved atmospheric model is a little more complex than that of the eponymous shallow pot due to the fact that the height separation between the surface and the rigid lid in a numerical model will typically be greater than a density scale height. As such, the mass distribution will be significantly biased towards the bottom of the model and so we need to establish how much of an error we would expect for such an idealized "deep" atmosphere (where "deep" means that the model depth, $z \geq H$, where H is the pressure scale height that is the same as density scale height for an isothermal atmosphere).

We can estimate the error due to "the pressure cooker effect" by examining how the diagnosed surface pressure varies within a sealed, vertically resolved, one-dimensional column. We take an initial total mass of atmosphere in the column, a specified mesh of computational levels in the model up to a maximum height of the model top at $z_{top}$, and an initial atmospheric temperature. We assume the atmosphere is isothermal for ease of analysis, but the result does not depend upon this assumption. We can calculate the density structure of the initial model atmosphere and from this we are then able to obtain the pressure at the model top, $p_{top}$. The model top pressure is a direct measure of how much of the atmospheric mass is above the initial model top and since the model top is rigid and the model is mass conserving, this $p_{top}$ remains constant for the model at all later times. For the initial structure, we can use the density ρ and temperature $T$ at each layer center to diagnose the corresponding pressure from the ideal gas law $p = \rho RT$. Finally, the surface pressure can be diagnosed from the hypsometric relationship

$$p_s = p_c \exp\left(\frac{g\Delta z_s}{RT_c}\right) \tag{1}$$

knowing the thickness between the surface and the lowest layer center $\Delta z_s$, the air temperature $T_c$, and the pressure $p_c$ at the center of this layer. For the initial state, this estimate of surface pressure will agree with one calculated by vertically integrating the density structure from the top down via hydrostatic equilibrium relation.

For the initial reference temperature, we will have a properly diagnosed pressure structure in which the surface pressure determined from multiplying the total atmospheric mass per unit area by gravity will equal the diagnosed surface pressure from the hydrostatic extrapolation from the lowest layer center (*i.e.*, $\int_{p_s}^{p_t} dp = -g \int_{z_s}^{z_t} \frac{p}{RT} dz$, with subscript "*s*" for surface and superscript "*t*" for top respectively); we have simply discretized a hydrostatic atmospheric profile. The potential error due to "the pressure cooker effect" will be felt as we change the temperature of the discretized "model" atmosphere. For example, if the isothermal temperature is increased, the density scale height (same as pressure scale height $H = \frac{RT}{g}$ in an isothermal atmosphere) will increase and thus the column will expand. Near the surface, the density will decrease, but this is compensated by the slower decrease of density with height such that above the lowest scale height the density will increase when the air temperature is increased. At the model top, the density will have a larger value than in the cooler reference case. In a real atmosphere (or a ventilated model), this would correspond to some of the atmospheric mass being pushed upwards out of the domain, since a larger density and temperature would correspond to a higher pressure at the model top, and this corresponds to a larger implied mass of air above the model top. In a pressure-related vertical coordinate model, it would correspond to the model top being deflected upwards in height. However, because of the rigid lid this cannot happen with a heigh-related coordinate model. Since the model with a rigid lid conserves mass, the total mass in the model domain (from the surface to the lid) remains fixed, even though a real atmosphere can freely redistribute its mass. At the warmer temperature, the real atmosphere would now have less mass over this same height range, *i.e.*, some mass would have moved above the level corresponding to $z_{top}$ in a real atmosphere. This "extra" mass now has to be accommodated within the model domain and is hydrostatically redistributed. This means that the density throughout the model column is now somewhat larger than it would be in the real (ventilated or unbound) atmosphere. Since the density is larger, the pressure is also everywhere higher in the model than in the real atmosphere at the same temperature according to the ideal gas law. This spurious extra pressure, when diagnosed at the surface, is called "the pressure cooker effect." It can be physically thought of as being the sum of the original surface pressure that is simply due to the gravitational weight of the air and the reaction pressure against the rigid lid due to warming.

It is straightforward to use the discussion in the prior paragraph to construct a simple quantitative model of "the pressure cooker effect." For application to Mars, we use a gas constant, $R = 191.84 \, Jkg^{-1}K^{-1}$, gravitational acceleration, $g = 3.727 ms^{-2}$, a reference surface pressure, $p_\circ = 600$ Pa, and a reference temperature of $T_\circ = 180$ K. We initially determine a reference hydrostatic atmosphere for these conditions discretized onto a collection of uniform height meshes with model depths, *i.e.*, $z_{top}$= 25, 50, 75, and 100 km, and using 10,000 layers (an ultra-fine resolution is used so as not to introduce noticeable errors associated with linear interpolation over the exponential density profiles). We then recalculate the density profiles and diagnose the surface pressures assuming a rigid lid for a range of temperatures from 120-240 K at 5 K intervals.

The dependence of the surface pressure error upon the model top height, quantified relative to the known correct surface pressure of 600 Pa and for an initial 180 K air temperature, is shown in Figure 3. For a model top at 25 km (Fig. 3a), the error in surface pressure is just over a 5% surface pressure deficit at 120 K to over 7% surface pressure excess at 240 K. These percentages correspond to about 30-45 Pa. However, as the model top is increased, the size of "the pressure cooker effect" error drops dramatically. For a model top at 100 km (Fig. 3d), the error is less than 0.03% (and less than 0.005% for the 120-200 K range, corresponding to only about 0.03 Pa). The reason that model top height is so important is because of the exponential hydrostatic density fall-off with height. When the model top is roughly ten times the scale height, so little of the atmospheric mass is near the top that even dramatic differences in the true versus capped-model density profiles do not correspond to much total error in mass predicted to be above the model top. The same analysis can be undertaken for changes in the atmospheric mass associated with the Martian seasonal cycle and with similar results for a 100 km model top. Thus, while the model density distribution at great height will be increasingly incorrect as the mean air temperature deviates from the initial state, the influence on lower and middle atmospheric dynamics will be minimal. Great skepticism need be applied to any dynamics near the top of any numerical model, due to the necessarily artificial treatment of the domain edge (a sharp edge which obviously does not, in reality, exist), though the rigid lid provides different, but not worse, concerns than those of the impermeable but mobile lids in pressure-based models. Thus, so long as the model domain is constructed to be deep enough for a given planetary atmosphere, then the "pressure cooker effect" can easily be reduced to a negligible concern. Specifically, the choice of model domain depth needs to be estimated based on the planetary scale height and the known range of mass-weighted, vertical mean air temperature.

**3.1 "The Pressure Cooker Effect" in MarsMPAS Simulations**

The performance of the MarsMPAS model with regards to "the pressure cooker effect" can be assessed directly from model output. Since the total mass of the atmosphere is known, the deviation of the area weighted surface pressure from the global air mass (per unit area multiplied by gravity) can be used to assess the error associated with this effect. In Figure 4, we show output from two separate MarsMPAS runs, the seasonal $CO_2$ cycle is active in one simulation and passive in the other. Figure 4a shows that in the full model, the error in diagnosed surface pressure is never greater than about 0.015%. This corresponds to an air temperature variation of just under 20 K (Fig. 4c). Comparison with Fig. 3d, for a model top of 100 km, we see that the error is significantly larger (by about a factor of 10) than the expected error (though note that it still only corresponds to about 0.1 Pa absolute error in surface pressure). Indeed, comparison of Fig. 4a with 4e suggests that error in the conversion processes between $CO_2$ ice and gas during the seasonal cycle likely explains most of the error and that this completely dominates "the pressure cooker effect" in the full MarsMPAS simulation. If the seasonal $CO_2$ cycle is deactivated, then Fig. 4b shows that the diagnosed surface pressure error is, indeed, reduced by the expected factor of 10 (note the y-axis scale change between Fig. 4a and 4b). Hence, as expected based on theory and shown in Fig. 3, when the model top is placed at ~100 km or greater for the current Mars atmosphere, "the pressure cooker effect" will have a negligible impact on Mars simulations.

## 4. Computational Performance of planetMPAS Compared to planetWRF

We assess the computational performance of MPAS in order to fully understand its capability as a cross-scale global circulation model. The results suggest that MPAS has superior performance towards fine grid resolution (0.5 degree or smaller). Two sets of experiments are performed: the comparison among MPAS, WRF and MITgcm and the scalability within MPAS itself. The detailed configurations and performance for all GCMs are listed in Table 1. All three models use the same set of physics parameterizations to allow performance evaluation focusing on the dynamical cores. MITgcm suffered numerical instabilities at horizontal resolutions of 1 degree and smaller without fine tuning the numerical filers, thus only 4-degree and 2-degree resolutions are shown here. WRF simulations marked with an asterisk (*) are probably overoptimistic for the timestep for numerical stability for simulations lasting more than 1 day from spin-up; a much shorter timestep is likely required (a prior, stable, spun-up 0.2-degree global WRF simulation required a 0.25 s timestep).

MPAS surpasses the performance of WRF at horizontal resolutions less than 1 degree due to the homogeneity of the horizontal grids. The model time step used by MPAS scales well with the model resolution, *i.e.*, the model time step decreases proportionally with increasing horizontal resolution (1, 0.5, and 0.25 degree resolutions in Table 1), while WRF requires much smaller time steps when the horizontal resolution decreases in order to maintain numerical stability in the polar region (the polar convergence problem). Additionally, the number of grid points used by WRF for a given resolution is much larger than that needed by MPAS, *e.g.*, WRF uses nearly 60% more horizontal grid points for 0.25-degree resolution. The combination of smaller time step and larger number of grid points leads to a more than 10 time slower performance of WRF compared to MPAS at 0.25-degree horizontal spacing.

MPAS is the least efficient code at relatively low resolutions, *e.g.*, 1, 2, and 4 degree grid spacing. MPAS is slower than WRF and MITgcm for calculations in the dynamical core that includes both the governing equations of fluid dynamics and the numerical filters. There are two reasons for the slower performance of MPAS. First, the MPAS dynamical core solves full 3D nonhydrostatic equations, *i.e.*, the vertical momentum equation is 3D in MPAS while it is assumed to be in hydrostatic balance for MITgcm. Furthermore, MPAS solves the 3D density field for the fully compressible atmosphere, while MITgcm solves the 2D pressure field (the surface pressure anomaly in an incompressible atmosphere). Second, MPAS carries 14 tracers by default despite the fact that all simulations in the benchmark case only are required to calculate the advection of two tracers, argon and $CO_2$ ice. The additional tracers in MPAS can be useful for multi-phase and multi-momentum microphysics but they will also slightly impact the efficiency of simple tracer simulations since the redundant tracers cannot be switched off entirely without modifying the dynamical core.

Both MPAS and MITgcm are slower than WRF in terms of per grid point performance at mid-to-low resolutions, despite the fact that MPAS and MITgcm use fewer grid points for a given horizontal resolution. This is mainly because WRF uses the finite difference method (FDM) to solve the fluid dynamic equations as opposed to the finite volume method (FVM) used by MPAS and MITgcm. FDM is faster than FVM due to its simplicity in orthogonal grids but is also less accurate due to roundoff errors incurred when approximating solutions using low-order Taylor expansions (*e.g.*, Lauritzen *et al.* [2011]).

MPAS exhibits excellent scalability with the number of processors for paralleling computing. Table 2 shows that the simulation speed increases with increasing number of processors proportionally, *e.g.*, the total wall clock used per day decreases by a factor of two when the number of processors doubles. However, the performance gain from increasing number of processors will be flattened when the number of ghost cells for halo exchange becomes comparable to the number of grid cells for each sub-domain (as with MITgcm, not shown). These findings are in line with the results from previous MPAS scaling experiments (Heinzeller *et al.* [2016]).

## 5. MarsMPAS

MarsMPAS uses a set of physics suite that closely follows those described in detail by Richardson and Newman [2018] (see their Appendix B). For radiative transfer, we use the k-distribution model (KDM) including the radiative effects of $CO_2$ gas and dust aerosols. The dust distributions are either prescribed following a modified version of the Mars Climate Database (MCD) formulation or predicted by the lifted dust due to near-surface wind stress, dust devil or surface injection. The surface energy balance and ground temperature are modeled using a 12-layer subsurface heat diffusion and surface energy balance model with TES albedo and thermal inertia maps. The YSU planetary boundary layer (PBL) scheme is used to parameterize the convective processes as diffusive mixing of momentum, heat and tracers. The seasonal $CO_2$ exchange between the surface (*i.e.*, the polar ice caps) and atmosphere as well as the formation of $CO_2$ ice clouds are included. Argon enrichment or depletion is calculated based on the amount of $CO_2$ gas added or removed from a grid cell due to $CO_2$ ice formation or sublimation.

## 5.1 MarsMPAS Modifications

### 5.1.1 Topography and Surface Properties

Surface property initialization for MarsMPAS generally follows that of MarsWRF but includes some upgrades. Like MarsWRF, surface albedo and thermal inertia are initially interpolated to MPAS mesh from spacecraft observations as initial conditions and are then updated during the simulations whenever they are modified by deposition or removal of surface $CO_2$ ice. Topography maps including elevation, surface slope magnitudes and orientation, and surface roughness are interpolated to the MPAS mesh (Figure 2). Additionally, the orographic standard deviation, asymmetry, and convexity are calculated from the topography map and projected onto the model grid as needed by the orographic gravity drag parameterization, which has yet to be implemented in MarsWRF.

### 5.1.2 $CO_2$ cycle

Taking advantage of the MPAS nonhydrostatic, fully-compressible dynamical core where atmospheric density is prognostic, $CO_2$ condensation and sublimation (which cases removal or addition of atmospheric gas) and their associated enrichment and dilution of non-condensable tracers are now handled within each grid voxel. This is different from pressure-based coordinate models, where air mass changes are captured in the prognostic surface pressure and hence immediately influence the whole column. At the same time, $CO_2$ ice particles advect both horizontally and vertically as other tracers on grid scale, and sediment with a fall velocity calculated from an assumed mean particle size, which is estimated from a typical condensation nuclei number density (CCN) of $10^5 kg^{-1}$ (Forget *et al.* [2013]). The restriction of pressure-based coordinate models is usually combined with an assumption that $CO_2$ condensates all deposit to the ground instantaneously and surface pressure is adjusted to reflect the pressure drop due to $CO_2$ condensation. Conversely, $CO_2$ ice only sublimates from the ground and the sublimation increases surface pressure accordingly. An alternative method to approximate local $CO_2$ condensation within the atmosphere, and its associated local enrichment/dilution of non-condensable tracers, is to rescale the vertical pressure coordinate from the top down to mimic the local subsidence and uplift of atmosphere due to $CO_2$ cycle (such as the scheme used in Mars MITgcm, Lian *et al.* [2012]). However, the rescaling of the pressure coordinate occurs instantaneously and advection of $CO_2$ ice particles is prohibited.

We further limit the fraction of gaseous $CO_2$ that can be condensed when the temperature is below frost temperature. This approach mimics the scenario where clouds only take a fraction of a grid box (*e.g.*, Smith *et al.* [1999]) and the condensation process takes a finite amount of time, therefore allowing $CO_2$ supersaturation in the rest of this grid box, *e.g.*, the atmospheric temperature can be sustained below the frost temperature (Colaprete *et al*. [2008]). From a numerical perspective, limiting the amount of $CO_2$ condensation per time step avoids "shocking" the atmospheric pressure by "sipping" (rather than "gulping") $CO_2$ gas out of the atmosphere, therefore enhancing numerical stability. Although convection driven by $CO_2$ condensation/sublimation is explicitly resolved at the grid scale by MPAS nonhydrostatic dynamical core, a more realistic approach with subgrid-scale convective mixing can be achieved using a $CO_2$ convection model with detailed microphysics similar to that used in NASA Ames MGCM (Colaprete *et al*. [2008]). We defer this as a future model development given the complicity involved with mass conservation.

### 5.1.3 Tracer Feedback

MPAS treats the tracer mass explicitly in the thermodynamic and momentum equations (Skamarock *et al.* [2012]). A "moist" density (the total density of dry air and tracers) is included in the pressure gradient terms of horizontal and vertical momentum equations, and a "moist" temperature replaces the dry potential temperature in thermodynamic equation. This is an alternative form of virtual temperature that is widely used in numerical models based on primitive equations. In the Martian atmosphere, non-condensable tracers such as argon and nitrogen cannot be simply treated as passive tracers. During the $CO_2$ cycle, local $CO_2$ condensation creates a vacuum effect that causes a discontinuity of air density — a potential source of acoustic waves. A moist density can help to suppress the effect by including the non-condensable tracer masses left behind at the locations of $CO_2$ condensation. For instance, the amount of argon or nitrogen mass is notable near the surface during polar winters. The mass mixing ratio of these non-condensables can occasionally exceeds 50% at some locations. Including the mass effect of the tracers will reduce the vertical and lateral pressure gradients at these locations. The default tracer contributing to moist density and moist temperature in the MPAS dynamical core is water vapor. We added a second tracer to represent non-condensables such as argon for Mars or condensables such as ammonia for Jupiter.

Partial pressures of tracers also contribute to radiative transfer and $CO_2$ condensation. In MarsMPAS, the KDM radiative transfer incorporates the radiative properties of $CO_2$ gas, water vapor and dust aerosols. Tracers such as argon and nitrogen are assumed to be radiatively inactive. The modification to $CO_2$ condensation is simple, as it uses $CO_2$ partial pressure ("dry" air pressure) to calculate the frost temperature of $CO_2$ (Wood and Paige [1992]).

### 5.1.4 Corrections of Energy and Momentum

The $CO_2$ cycle complicates the conservation properties of momentum and thermal energy as $CO_2$ is the major component of the air. MPAS solves a conservative form of continuity, momentum and thermal dynamic equations. However, the conservation of energy and momentum only applies to the gas phase, meaning that the temperature and wind speed in a control volume needs to increase to compensate the removal of air mass. Physically the $CO_2$ condensed out of the air still carries thermal energy (*e.g.*, at frost temperature) and momentum (*e.g.*, at wind speed before condensation), while this portion of $CO_2$ is unaccounted for the total energy and momentum budget in the current dynamical core. In order to mitigate this energy scaling effect without modifying the dynamical core, we treat $CO_2$ ice as an energy reservoir that stores thermal energy removed from the air. In addition to the latent heating effect, a cooling rate derived from conservation of energy during $CO_2$ condensation is applied to the environmental air to maintain the total energy budget (air and ice) unchanged. A similar procedure also applies when $CO_2$ ice sublimates. The momentum is more difficult to correct since $CO_2$ ice does not carry momentum explicitly. Fortunately, most $CO_2$ condensation occurs at ground level. Therefore, adding momentum removed by $CO_2$ condensation to the uncondensed portion of $CO_2$ air at the lowest model level will compensate for the majority of the total momentum budget. Conversely, $CO_2$ sublimated from the surface will lead to a slowdown of the wind speed in the lowest model layer. Lastly, non-condensable tracer mass is conserved by adjusting the mass mixing ratio when $CO_2$ air mass changes due to condensation or sublimation.

### 5.1.5 Orographic Wave Drag Parameterization

Internal gravity waves are ubiquitous features in the Martian atmosphere. These waves have been observed by multiple spacecraft such as Mars Global Surveyor (MGS) and Mars Atmospheric and Volatile EvolutioN (MAVEN) throughout the lower (2-50 km) and middle atmosphere (50-100 km), all the way to the thermosphere / ionosphere (100-250 km) (e.g., Banfield *et al*. [2000];

Withers [2006]; Creasey *et al*. [2006a, 2006b]; Bougher *et al*. [2017]). These waves could be generated by terrain variation, breaking thermal tides, or convection. Among these types of gravity waves, orographically generated gravity waves are relatively easier to parameterize in global circulation models since their correlations with terrain features are more predictable than transient waves.

The energy and momentum transported by vertically propagating gravity waves can significantly affect the local dynamics and thermal structure of an atmosphere. Wave breaking and dissipation can impact the dynamical structures of the atmosphere via turbulent mixing of heat, momentum and chemical constituents, notably in the stratosphere and above (*e.g*., Fritts and Alexander [2003]). On Mars, gravity waves likely have a significant impact on the thermal and wind structures of the middle atmosphere and thermosphere, as well as on the atmospheric constituent distributions. Collins *et al*. [1997] implemented a gravity wave drag scheme devised by Palmer et al. [1986], Baines and Palmer [1990] and Lott and Miller [1997] to represent breaking of orographic waves in their global circulation model. With the scheme active, they were able to produce winter polar warmings in close agreement with temperatures retrieved from Mariner 9. Similarly, Medvedev *et al*. [2015] implemented a gravity wave drag parameterization in both the LMD and Max Planck Institute (MPI) Mars GCMs, and they found that the divergence of energy flux induced by wave dissipation (due to saturation/breaking and viscous/diffusive damping) could produce cooling rates that explain discrepancies between GCM model predictions and spacecraft observations (*e.g*., from Mare Express SPICAM) in the lower thermosphere, while the wave dissipation could also induce downdrafts of polar, high-altitude air necessary to increase the polar warming in the mesosphere. Numerical studies also suggested that the orographically generated gravity waves contribute significant amount of turbulent energy and wind drag to atmospheric circulations (*e.g.*, Kuroda *et al*. [2016]).

Here we implement the orographic gravity wave drag parameterization within the WRF and MPAS physics package. This parameterization is based on the scheme developed by Kim and Arakawa [1995] (hereafter, KA95), where the reference-level drag (drag produced by unstable wave motions when Richardson number exceeds a prescribed critical value, typically 0.25) is formulated to include the effect of breaking hydrostatic and non-hydrostatic gravity waves near a critical layer where wave motion becomes unstable. The major difference between this scheme and those adopted by Collins et al. [1997] is the treatment of low-level wave breaking due to resonant amplification of nonhydrostatic waves trapped to the downstream (lee side) of mountains. KA95 extensively tested the low-level wave breaking using a 2D nonhydrostatic anelastic model and found that the drag induced by low-level wave breaking could be better parameterized compared to previous work, *e.g.*, Pierrehumbert [1986].

The reference-level drag $\tau_0$ in the KA95 scheme is expressed as:

$$\tau_0 = E \frac{m'}{\Delta x} \frac{\rho_0 U_0^3}{N_0} G'$$

Where $\Delta x$ is the model grid size; $\rho_0$, $U_0$ and $N_0$ are the density, square root mean wind, and buoyancy frequency at the reference level that is at least two times the standard deviation above the surface; $E$, $m'$ and $G'$ are an enhancement factor, number of mountains, and a flux function that depend on the terrain features including standard variation (SD), orographic asymmetry (OA, representing the placement of a barrier relative to the wind direction in a given domain), orographic convexity (OC, a measure of sharpness of the barrier) as well as orientation of barrier relative to the wind directions. The detailed definitions and numerical expressions for these statistical orographic parameters can be found in KA95 and Kim and Doyle [2005]. In MarsMPAS, SD, OA and OC are calculated using Mars MOLA map with 1/32° resolution, which is sufficient for the nominal model resolutions of 2° or 4°.

Two major tunable parameters $C_E$ and $C_G$ are the constants related to the enhancement factor $E$ and flux function $G'$ via relations:

$$E = (OA + 2)^{C_E Fr_0 / Fr_c}$$

$$G' = \frac{Fr_0^2}{Fr_0^2 + C_G OC^{-1}}$$

Where $Fr_0$ and $Fr_c$ are the Froude number at the reference level and a prescribed critical Froude number respectively. In MarsMPAS, we use $C_E = 0.2$ (vs 0.8 for Earth) and $C_G = 7.$ (vs 0.5 for Earth) to maintain the reference-level drag at a reasonable strength that will not excessively modify the low-level circulation patterns compared to the case without gravity wave drag parameterization. We choose large $C_G$ to make it consistent with that of the Earth conditions where the ratio $C_G/OC \approx 1$ because the mean convexity on Mars is larger than that on Earth. The smaller $C_E = 0.2$ is chosen to reduce the enhancement factor directly since the standard deviation of terrain on Mars is also larger than that on Earth.

Once the reference-level drag is determined, the vertical structure of the drag above the reference level can be calculated via wave saturation theory (Lindzen [1981]). Currently the gravity wave drag parameterization only applies to momentum. However, wave breaking can also cause diffusive heating and cooling of the atmosphere directly (*e.g*., Schoeberl *et al.* [1983]). This process will be considered in a future version of the parameterization. A detailed tunning of the parameters in the KA95 scheme under the Martian atmospheric condition would also be conducted in a separate paper.

### 5.1.6 Two-Moment Dust Scheme

MarsMPAS, similar to MarsWRF, has several options to describe temporal and spatial variation of atmospheric dust loading. These options include prescribed dust distributions such as the MCD-MGS and Conrath profiles that are functions of latitude, pressure and solar longitude (Richardson *et al*. [2007]); dust lifted or injected from surface with one or two prescribed particle sizes (Newman *et al.* [2002]); and two-moment dust distributions where dust mass mixing ratio and number concentration are prognostic via advection and microphysical processes (Lee *et al.* [2018]; Morrison and Gettelman [2008]).

The two-moment dust scheme allows the evolving dust distribution to interact with solar and thermal radiative transfer to yield atmospheric heating rates. It also allows dust to interact with water ice particle distributions via ice coating forming on the dust particle (*i.e.,* water using the dust as cloud nuclei). MarsMPAS currently implements two-moment dust only, *i.e.*, the 'dry' case described in Lee *et al.* [2018]. In this scheme, dust particles initially lifted from the surface are characterized by a Gamma distribution with a fixed effective radius, where the distribution function establishes the relationship between the dust mass mixing ratio and number concentration. Currently MarsMPAS supports two types of surface dust sources: 1) dust lifted by physical processes, such as dust devils and near-surface wind stress (Newman *et al.* [2002]), and 2) dust injected from the surface by comparing the column-integrated dust opacity in MarsMPAS to those derived from spacecraft observations during MY24-MY34 (Montabone *et al.* [2015]). Dust particles in the atmosphere can be transported between voxels via advection, sub-grid scale mixing, and sedimentation.

**5.1.7 Variable Gravity**

PlanetMPAS has the option to use a constant gravity for the entire atmosphere or variable gravity as a function of radial distance from the center of the planet. It is traditional to assume that atmospheres are thin compared to the radii of planets. However, the variability of gravity can no longer be ignored when studying extended atmospheres, such as thermosphere on Mars and Titan and deep atmospheres on gas giant planets. Modifying MPAS to use variable gravity is straightforward if the variability is assumed to be small, that is, $|\frac{1}{g}\frac{\partial g}{\partial r}| \ll \frac{1}{H}$, where $g$ is the gravity, $r$ is the radial distance and $H$ is the density scale height. In this case, the gravity in the MPAS vertical momentum equation (see Equation 4 in Skamarock *et al.* [2012]) can be simply replaced by $g(z, \phi, \theta) = \frac{GM}{(a+z)^2}$, where $\phi$ is the azimuthal angle, $\theta$ is the polar angle, $G$ is the gravitational constant, $a$ is the planet radius and $z$ is the distance above the geoid. The variable gravity is initialized as a 3D static field since MPAS uses height coordinate, which does not vary with time unlike the surface pressure anomaly in pressure coordinates. Similarly, the constant gravity in physics parameterizations can be replaced by the variable gravity defined above.

The recently-developed deep-atmosphere component of MPAS has also implemented variable gravity similar to the one we use for planetMPAS. Additionally, it considers some Coriolis force and curvature terms as well as the variability of the gas constant and heat capacity ignored in a shallow-atmosphere model (Skamarock *et al*. [2021]). However, this version of MPAS has not been released to public during the development of planetMPAS, but we expect to upgrade planetMPAS with the new MPAS dynamical core soon after it is released.

**5.1.8 Rayleigh Damping and Vertical Smagorinsky Viscosity**

MPAS uses Rayleigh damping of vertical velocity to suppress the growth of vertically propagating waves generated either physically or numerically. The starting point of this damping (hereafter, the damping altitude) is a fixed height above sea level in the original MPAS setup. However, we find that this method may overly damp the vertical motion or allow the growth of waves to destabilize the atmosphere because of steep terrain variations. For instance, the damping altitude needs to be about 30km or 35km above Olympus Mons to suppress waves associated with sharp terrain. The convectively generated waves (due to $CO_2$ convection) in the winter polar regions would be under-damped if we were to use a fixed damping altitude 30km above Olympus Mons (about 45km above the sea level) because these convectively generated waves usually start around 20 to 30km (sometimes even lower) above the zero datum ("sea level") in our simulations. Similarly, orographic waves over Olympus Mons would be over-damped if we lower the damping altitude to suppress the convectively generated waves. Therefore, we modified the damping altitude to be terrain-following, meaning that the reference height of damping altitude is the terrain elevation. Tests based on the modified damping altitude suggest that the modification do not change the overall dynamics.

We also implement a second-order vertical Smagorinsky viscosity that is estimated from horizontal wind shear and a prescribed vertical mixing length, *e.g.,* local vertical grid spacing (Smagorinsky [1993]). This vertical Smagorinsky viscosity replaces the fixed vertical eddy viscosity in the original MPAS setup to work with the second-order horizontal Smagorinsky viscosity, which is implemented by both WRF and MPAS to suppress numerical instabilities. Here we use both the horizontal and vertical Smagorinsky viscosities to achieve better numerical stability as well as maintain consistency of viscous dissipation of 3D flows (Smagorinsky [1993]). This modification further provides a rule of thumb on the tuning of YSU PBL scheme in the free atmosphere.

## 5.2 MarsMPAS Model Setup

We have chosen a model setup that is largely consistent with both MarsWRF and Mars MITgcm for better inter-model comparison. The initial atmosphere is motionless and isothermal at 300K, and contains a non-condensable tracer gas with a typical mass mixing ratio of 0.0145 that is uniformly distributed everywhere. The nominal horizontal meshes used in the MarsMPAS cases have 10,242 and 2,562 horizontal grid points nearly uniformly spaced, which corresponds to roughly 2 degrees (120km) and 4 degrees (240km) grid spacing. For a given horizontal resolution, the number of grid points used by MPAS is about 35% less than that in longitude-latitude grid due to the uniformity of MPAS mesh. Higher horizontal resolutions such as 0.5-degree and 0.25-degree fixed resolutions and 0.75-0.25 degree variable resolution are also used to test the performance of model scalability (see Section 4). Vertically there are 45 atmospheric layers that extend from the surface to 120km above datum, and there are 12 sub-surface layers that conduct heat from atmosphere into the regolith. Additionally, we tested 100 vertical atmospheric layers to examine the impact of resolvable vertically propagating waves on the model. The result, consistent with the findings by Skamarock *et al.* [2019], shows that 100 vertical layers indeed reduces the spurious oscillation of temperature near the model top. However, we use 45 layers for the rest of the simulations to reduce computational cost. Instead, a sponge layer near the model top is applied to eliminate wave oscillations there. The model time step is 120 seconds for both 2-degree and 4-degree resolutions and it is limited by the vertical CFL. The radiative transfer is updated every 12 minutes. The $CO_2$ condensation/sublimation, boundary layer parameterization, and surface and subsurface schemes run every model time step.

### 5.3 MarsMPAS Model Results

Initial tests with MarsMPAS show that the model produces seasonal variations of large-scale features comparable to or better than previous Mars GCMs including MarsWRF and Mars

MITgcm, such as global mass conservation, Viking lander pressure cycles (mostly a result of the $CO_2$ cycle), zonal mean temperature, and polar argon enrichment. These aspects have become benchmarks for a model's ability to represent real Martian atmosphere, *i.e.*, good matches between model results and the observations are the prerequisites for more complex modeling of the Martian atmosphere. In the following sections, all cases were run using the prescribed "MCD MGS" dust scenario unless otherwise specified.

**5.3.1 Mass Conservation and $CO_2$ Cycle**

MarsMPAS shows excellent mass conservation properties largely due to its finite volume dynamical core. We have also ensured that the parameterized physics processes associated with the $CO_2$ cycle, such as $CO_2$ condensation/sublimation (both in air and on surface) and $CO_2$ ice particle transport (including advection and sedimentation) are all mass-conserving. Figure 5a and 5b show the example of the zonal mean mass mixing ratios of $CO_2$ ice in air at $L_s = 90°$ and $L_s = 270°$ respectively. Most $CO_2$ condensation occurs near the surface (*i.e.*, within 5 km above the ground) during southern winter while it occurs at higher altitudes (*i.e.*, 5 km to 15 km) during northern winter. $CO_2$ also condenses at relatively high altitude (~ 25 km) near equator (Fig. 5a), which may be caused by the adiabatic cooling of air flow over Tharsis Montes.

Figure 6 shows the total global atmospheric $CO_2$ air mass, surface and atmospheric $CO_2$ ice mass, and Argon mass as a function of solar longitude. MarsMPAS is initialized with a total atmospheric mass of about $2.85 \times 10^{16}$ kg, which is chosen to reproduce Viking Lander pressure cycles (Fig.

7). Similar to the predictions by other Mars GCMs, more than 25% of total air mass can be condensed during southern winter. Both total $CO_2$ and non-condensable tracer (Argon) masses show little variation over a Martian year (*e.g.*, a change of ~100 kg compared to the total Argon mass of $4.14 \times 10^{14}$ kg during a Martian year). This variation is mostly a result of accumulated machine error over a long integration interval (*e.g.*, a few Martian years).

With some tuning of surface albedo and emissivity (Guo *et al*. [2009]), MarsMPAS is able to reproduce the seasonal variation of $CO_2$ pressure cycles measured at Viking Lander 1 and Viking Land 2 sites. Figure 7 shows the simulated pressure curves compared to the best fits of pressure measurements at both Viking Lander locations. Overall, the simulated and the observed pressure variations have good match, with only small difference (within 3%-5%) existing during autumn equinox and northern winter. Such differences are also unlikely to have a significant impact on the overall dynamics, nor tracer enrichment and depletion.

### 5.3.2 Seasonal Temperature Variations

The seasonal and spatial variations of zonal mean temperature in MarsMPAS are comparable to those retrieved from MCS measurements and at least comparable to other Mars GCM predictions (*e.g.*, McDunn *et al*. [2013]; McCleese *et al*. [2010]; Forget *et al*. [1999]; Richardson *et al*. [2007]; Lian *et al*. [2012]; Brecht *et al*. [2019]; Bertrand *et al*. [2020]). Figure 8 shows the zonal mean temperature at spring ($L_s = 0°$), northern summer ($L_s = 90°$), autumn ($L_s = 180°$) and northern winter ($L_s = 270°$). With orographic gravity wave drag (GWDO) switched on, MarsMPAS is able to reproduce the magnitude of polar warming comparable to the observed values (*e.g.*, McCleese *et al.* [2010]). Even without GWDO, the strength of the polar warming in MarsMPAS exceeds that in both MarsWRF and MIT Mars GCM (Lian *et al.* [2012]), likely due to stronger vertical motions near the poles permitted by the MarsMPAS non-hydrostatic dynamical core.

Figures 9 and 10 show the differences of temperature and zonal wind between the cases with and without GWDO. During all seasons, GWDO slightly slows down the zonal wind in the polar regions above 60 km (Fig. 10), thus permitting a weak poleward flow near 60 km. The downdraft of this poleward flow leads to adiabatic warming at altitudes between 40 km and 60 km (Fig. 9). The effect of GWDO on the polar warming is similar to the findings by Collins *et al.* [1997]. A more intuitive way of viewing the effect of GWDO is to compare the residual mean circulations (*i.e.*, Transformed Eulerian Mean circulations (Holton [2004])). Figure 11 shows the difference of stream functions associated with residual mean circulations between the cases with and without GWDO at $L_s = 90°$. The presence of GWDO induces an anti-clockwise circulation between 45 km and 65 km. The downward branch of this circulation is what causes the polar warming seen in Fig. 8b.

It is worth mentioning that treating the vertical component of Smagorinsky viscosity is necessary for the model to produce the proper thermal structure in the lower 40 km in MarsMPAS. The atmosphere would become too warm below 40km without the vertical Smagorinsky viscosity despite both MarsMPAS and MarsWRF utilizing the same horizontal Smagorinsky viscosity as a numerical filter. The discrepancy may be due to the differences in the dynamical cores. MarsWRF when run on global scales solves the 3D primitive equations that do not permit strong vertical motions and hence limits the flows mostly to 2D horizontal planes. Therefore, the horizontal component of the Smagorinsky viscosity is sufficient to represent the turbulent mixing of energy and momentum (mostly 2D) in MarsWRF. MarsMPAS, on the other hand, solves fully 3D non-hydrostatic equations that allow strong vertical motions (the vertical velocity can occasionally reach tens of meters per second during dusty seasons). The horizontal and vertical turbulent mixing of energy and momentum may become equally important when strong convection presents. Under this circumstance, both horizontal and vertical Smagorinsky viscosities need to be considered to represent 3D turbulent mixing properly.

### 5.3.3 Argon Enhancement

MarsMPAS is able to produce a maximum Argon enhancement factor (EF) of about 6.5, which is close to the GRS measurement of ~6 (Sprague *et al*. [2007]) and larger than the value (EF=4.75) predicted by the best-quality advection scheme in the MIT Mars GCM (Lian *et al*. [2012]). Figure 12 shows the EF as a function of solar longitude in both south and north polar regions (75° poleward) using the same method described in Lian *et al*. [2012]. Like those observed by GRS and predicted by previous Mars GCMs, south polar EF peaks at late southern winter, and reaches minimum during southern summer. North polar Argon, on the other hand, experiences very mild depletion and enrichment during southern winter and summer respectively.

The large south polar EF produced by MarsMPAS is perhaps due to better dynamical representation of the polar vortex than that in the MIT Mars GCM since a strong polar vortex can behave as a barrier to prevent meridional mixing of tracers between the polar region and mid-to-low latitudes (*e.g.*, Mizuta and Yoden [2001]). Unlike the MIT Mars GCM that has many choices of advection schemes, MarsMPAS only has a third-order upwind, positive-definite advection scheme that actually produced very weak EF (~2.5) in the MIT Mars GCM. This leads to our speculation on how model-to-model differences may affect the detailed dynamics of the polar vortex. Although the polar vortex behaves as a barrier, it is not a perfect barrier. Materials can still leak out of the polar vortex due to mixing processes across the polar vortex edge. A good quality tracer advection scheme can suppress the artificially diffusive transport of tracers from the polar vortex interior to its outer edge due to errors resulting from numerical approximation of advection, but physically it all comes down to the model's ability to realistically represent the polar vortex and the turbulent mixing processes within. The density-based $CO_2$ condensation in MarsMPAS may better represent the condensation flow that affects the strength of the polar vortex. At the same time, the Smagorinsky viscosity used by MarsMPAS is more physical than the Shapiro filter used by the MIT Mars GCM when suppressing grid-scale numerical noise, therefore better representing turbulent mixing processes within the polar vortex.

The peak value of EF is dependent on the dust forcing in MarsMPAS. We varied the dust forcing strength by increasing/reducing either the dust opacity (prescribed dust scenario) or the surface dust injection rate (interactive dust scenario) and found that EF is positively correlated with the dust forcing. This again suggests that EF is affected by the strength of the polar vortex since stronger dust forcing drives stronger poleward flow (thus stronger circumpolar zonal flow) and vice versa. Nonetheless, analysis on potential vorticity [*e.g.*, Toigo *et al.* 2017] in the polar region will be required to quantify the correlation between EF and the polar vortex.

**5.3.4 Interactive Dust**

Initial tests with fully-interactive dust show that both physically lifted dust and observationally injected dust scenarios can produce seasonal and spatial variations of temperature, $CO_2$ pressure cycle and Argon enrichment similar to the results from the prescribed dust scenario. The two-moment dust scheme has also been tuned to produce mean effective dust radii around 1.5 $\mu m$, which is close to the values derived from multiple spacecraft and rover observations of the Martian atmosphere (*e.g.*, Clancy *et al.* [2003]; Guzewich *et al.* [2014]; Chen-Chen *et al.* [2019]).

Figure 13 shows the density-scaled dust opacity and $CO_2$ ice opacity at selected solar longitudes ( $L_s = 90°$, $L_s = 150°$ and $L_s = 300°$ ) during northern spring and summer such that they can be compared to the MCS retrievals by Heavens *et al.* [2011]. Both dust opacity and $CO_2$ ice opacity are diurnally averaged for simplicity. The modeled dust opacity shares some similarities with the MCS observations at mid-to-low latitudes such as the range of values ( $10^{-6}\ m^2\ kg^{-1}$ to $3 \times 10^{-3}\ m^2\ kg^{-1}$) and the typical height the dust can reach (~ 10 Pa) during northern summer and late northern winter. Particularly, they are generally in good agreement near $L_s = 300°$ except in the north polar region (Fig. 13e). They also differ in several aspects. First, the modeled peak values of density-scaled dust opacity usually appear at lower altitudes compared to the MCS observations. This is clearly noticeable near $L_s = 150°$ (Fig. 13c) as the modeled dust has a lower top (20 Pa vs. 6 Pa in the MCS observations) and its peak value is located near 300-400 Pa that is much lower than the observed (~ 60 Pa). Second, the modeled dust and $CO_2$ ice opacity cannot capture the opaque features between 10 Pa and 100 Pa at either northern or southern winter pole. These features are thought to be $CO_2$ ice [*e.g.*, McCleese et al., 2010]. Figure 13b, 13d and 13f show the modeled $CO_2$ ice opacity at $L_s = 90°$, $L_s = 150°$ and $L_s = 300°$ respectively. Compared to the MCS retrievals (*e.g.*, McCleese *et al.* [2010]; Heavens *et al.* [2011]), it is apparent that the modeled $CO_2$ ice cannot explain the observed opacity at winter poles. This suggests that the modeled polar temperature between 10 Pa and 100 Pa may be too warm to produce sufficient $CO_2$ ice. Some fine-tuning of dust forcing would be needed to improve the thermal structures. This may also be due to the simple physics parameterization of $CO_2$ cycle in the current MarsMPAS, which treats the $CO_2$ ice as single sized particles (2 $\mu m$) without any interactions with dust particles. Provided a proper size distribution, $CO_2$ ice particles with smaller sizes would sediment slower and be transported to higher altitudes. The discrepancy in the dust opacity peak away from the surface may be associated with the model lacking a parameterization scheme that properly accounts for dust mixing in the PBL by radiatively active deep dust plumes [Rafkin, 2012; Colaïtis *et al.*, 2013; Wu *et al.,* 2021].

### 5.3.5 Variable-Resolution Simulations

One of the many useful features of MPAS is its ability to simulate a global domain with variable resolution, which provides a smoother transition from coarse resolution to fine resolution than two-way domain nesting. The downside, however, is that the time step of the entire simulated domain is limited by the smallest grid spacing. In comparison, *e.g.*, MarsWRF uses time steps adaptive to the grid sizes for different domains (*i.e.*, large time step for coarse grid and vice versa).

Variable resolution is particularly useful when examining local features such as pressure, temperature, and wind fields near lander/rover locations (*e.g.*, Newman *et al.* [2017]), as well as studying mesoscale dynamics such as gravity waves, local and regional dust storms, flows over areas of complex topography, *etc*. Figure 14 shows the simulated surface pressure in the vicinity of Gale Crater with uniform 2° resolution (Fig. 14a) and variable resolution ( 0.75° to 0.25°, Fig. 14b) at midnight near $L_s = 60°$. The fine grid resolution ( 0.25°, equivalent to about 14 km) can coarsely resolve Gale Crater which has a diameter of about 150 km and yields an average pressure of about 880 Pa near the Curiosity Rover location. This value is close to the pressure measured by REMS on board the Curiosity Rover during the similar time of MY33. The coarse grid resolution ( 2°), on the other hand, produces a pressure of about 720 Pa at the same location and time due to the poorly resolved terrain.

Figure 15 shows an example of gravity waves resolved by the fine resolution grid. Fig. 15a-15d are the time series of vertical velocity as a function of longitude and altitude at latitude $\phi = -4.49°$ (Gale Crater) near $L_s = 60°$. The output frequency of the time series is 30 minutes. A quick examination of wave features near 70 km above the ground suggests that this is likely a transient mesoscale wave with a phase speed of about $75 ms^{-1}$ and a wavelength of about 475 km. The variable-resolution simulations demonstrate that MarsMPAS is a capable tool to model waves in the Martian atmosphere across scales, *e.g.*, from orographic and convectively generated mesoscale gravity waves all the way to planetary scale waves such as global thermal tides.

**6. JupiterMPAS**

Jupiter is a gas giant planet that is vastly different from the terrestrial planets such as Earth and Mars. It has a primitive atmosphere primarily composed of hydrogen and helium without a solid surface. Jupiter exhibits banded cloud patterns that are relatively stable over hundreds of years (*e.g.*, Vasavada and Showman [2005]). These banded clouds are composed primarily of ammonia ice particles, and are closely correlated with east-west jets that extend thousands of kilometers into the interior of the planet, based on the analysis of the gravity field measured by Juno (Kaspi *et al.* [2018]). The formation of these banded clouds is still not well understood, as the internal heat flux (deep forcing), differential heating from solar radiation, and moist convection (shallow forcing) all play roles in producing the banded jets and equatorial superrotation (*e.g.*, Aurnou *et al.* [2008]; Kaspi *et al.* [2009]; Schneider and Liu [2009]; Lian and Showman [2008, 2010]). Recent analysis of Juno gravity measurements suggests that winds visible at the ammonia cloud top could extend 3,000-4,000 km towards the interior of the planet (*e.g.*, Kaspi *et al.* [2018]), which is consistent with the predictions from the deep convection models (*e.g.*, Kaspi *et al.* [2009]; Aurnou *et al.* [2008]). However, Juno Microwave Radiometer (MWR) observations of atmospheric tracer distributions indicate moist convection still dominates the weather layer dynamics and perhaps affects the circulations much deeper than the water cloud base (Ingersoll *et al.* [2017]; Bolton *et al.* [2017]).

Studying moist convection and its interactions with large-scale dynamics relies on numerical models such as GCMs and cloud resolving models (CRM, on horizontal scales of about 1 km and smaller). Jupiter GCMs have evolved from being idealized models to more complex models as increasing numbers of observations have become available as constraints, enabling more specific questions to be framed. Limited by computational resources and few observational constraints, early 3D GCMs (with a domain extending down to tens of bars) used simplified forcing mechanisms such as Newtonian heating/cooling to represent solar heating and radiative cooling of the atmosphere (*e.g.*, Lian and Showman [2008]), and large-scale condensation schemes to represent the (water) hydrological cycle (*e.g.*, Lian and Showman [2010]). Some GCMs used more realistic radiative transfer models such as two-stream semi-gray radiative transfer, but only for a dry atmosphere and with a vertical domain limited to a few bars (*e.g.*, Schneider and Liu [2009]; Spiga *et al*. [2020]). All these 3D GCMs have successfully produced jet patterns on Jupiter. There are also deep convection models (with domains extending down to thousands of bars) that can produce jet patterns on Jupiter (*e.g.*, Kaspi *et al*. [2009]; Aurnou *et al*. [2008]), but these models do not interact with the weather layer because it is computationally prohibitive to simulate both regions at once.

There have also been efforts to model the microphysical processes of various condensable species, such as water and ammonia, but these physics parameterizations have not been tested in 3D GCMs that are able to spontaneously produce jet patterns in the Jovian atmosphere (Palotai and Dowling [2008]). Young *et al*. [2019a, 2019b] developed a Jupiter GCM (*Jason*) based on the MITgcm dynamical core that was also adopted by Lian and Showman [2008, 2010]. They were able to produce banded jet patterns and equatorial superrotation using a semi-gray radiative transfer model and imposed internal heat flux in a dry atmosphere (Young *et al*. [2019a]), although their model predicted much weaker superrotation than that observed, with modeled winds of only 30 $ms^{-1}$ at the ammonia cloud top *vs*. 140 $ms^{-1}$ observed on Jupiter. The *Jason* model can simulate water and ammonia cycles with a moist convection parameterization that mixes heat and tracers (Zuchowski *et al*. [2009b, 2009c]), but Young *et al*. [2019b] only examined the passive tracer cycles and ignored the impact of latent heating/cooling on large-scale dynamics. Moreover, their moist convection scheme does not mix momentum associated with convective storms, which may be important for the Jovian atmosphere.

We have developed the JupiterMPAS model to represent weather layer dynamics in the Jovian atmosphere with complex physics parameterizations. The model physics represent a continuous development from Lian and Showman [2008, 2010], with significant improvements in the moist convection and cloud microphysics parameterizations. The variable-resolution grid in JupiterMPAS provides an opportunity to investigate convective processes at much higher resolution than in previous global models and hence the interaction of dynamics across meso- and synoptic (global) scales. Although the model cannot be run globally at a resolution high enough to be cloud resolving, it still acts as an intermediate solution to bridge the gap between planetary-scale motions (on order of $10^4$ km) and mesoscale convective motions (on order of hundreds of kilometers for Jupiter).

## 6.1 JupiterMPAS Physics Parameterizations

JupiterMPAS has the following sub-grid scale physics parameters established so far. There are four major components of physics parameterizations that have been implemented to simulate the moist weather layer and sub-cloud layer dynamics, where the sub-cloud layers refer to the dry convection layers below the water cloud base. In the sub-cloud layers, where the internal heat flux is the major driver of convection, a turbulent mixing model based on the Boulac scheme adapted to Jovian conditions is used to parameterize the turbulent mixing of horizontal momentum, temperature and tracers (Bougeault and Lacarrere [1989]). The Boulac scheme is based on a 1.5 order closure model, where the local subgrid-scale turbulent fluxes of momentum, heat and tracers are correlated to the vertical gradients of these components. The mixing length is estimated from the vertical scale at which turbulent kinetic energy (TKE) is fully consumed by the buoyancy. Therefore, the scheme is suitable for gas giant planets in general as it does not rely on empirical estimates of the mixing length (*e.g.*, an approach commonly used for Earth atmosphere). We have tuned the Boulac scheme to produce diffusion coefficients within the range estimated by Visscher et al., [2010].

The second component is a mass flux based moist convection scheme for the water cycle [Tiedtke, 1989]. This moist convection scheme includes parameterizations of shallow, mid-level and deep convections. For deep convection, a cloud base mass flux (closure) is estimated from the equilibrium between large-scale forcing (*e.g.*, large-scale circulations driven by solar forcing) that generates convective available energy (CAPE) and convection that reduces CAPE. This scenario is appropriate for the Jovian weather layer. Similarly, for shallow convection, the closure is based on the balance between boundary-layer forcing and convection. Shallow convection is primarily controlled by surface fluxes, which cannot be properly estimated since gas giants do not have a "surface". Mid-level convection is usually triggered by a large-scale ascent in a sufficiently moist environment, this may occur in the layers right above the dry convection deep layer, but we only consider deep convection in the current JupiterMPAS for simplicity. The Tiedtke moist convection scheme is applied to the water cycle in the Jovian atmosphere. It convectively mixes heat, momentum and tracers including water and ammonia vapor.

The third component is a simple cloud microphysics scheme based on Kessler [1969] and Palotai and Dowling [2008], which includes production of cloud water from ice melting, production of rain from cloud water and evaporation of rain and sedimentation of rain, evaporation of cloud water and rain, as well as sedimentation of rain and ice/snow. The production of cloud water from vapor is taken care of in the aforementioned Tiedtke moist convection scheme. Unlike the calculations in Palotai and Dowling [2008], terminal velocity is estimated by a more generalized method that works for a large range of temperatures, pressures, and Reynold numbers (Khvorostyanov and Curry [2002]). The microphysics also includes the production and evaporation of "mushballs" described by Guillot *et al*. [2020a, 2020b]. The mushballs form when adsorption of ammonia vapor by supercooled water ice particles occurs above 1.5 bars. They can grow to large sizes (*e.g.*, a few centimeters) by collecting ammonia and water condensates on the downfall, and evaporate in the region deeper than the water cloud base.

The fourth component is a moist convective adjustment scheme for the ammonia cycle. This scheme is based on a Simplified Betts-Miller (SBM) scheme (Frierson [2007]), which simplifies the original Betts-Miller scheme [Betts and Miller, 1986] by lowering the lifted-condensation level to surface and substituting an empirical reference moisture profile to relative humidity (80% typical in Frierson [2007]). Unlike the Tiedtke scheme, the SBM scheme only mixes heat and moisture. We chose the SBM scheme for the ammonia cycle because weather layer dynamics is mostly affected by water moist convection, and the SBM scheme is computationally much more efficient than the Tiedtke scheme. The choice is justified given that ammonia is significantly less abundant, and it condenses in a relatively higher, thinner region of the atmosphere compared to water vapor. We further assume that ammonia condensates fall instantly to the layer below the cloud base, where they sediment and evaporate. This is analogous to the terrestrial Bett-Miller treatment of condensates that fall directly to the surface.

Additional components of the physics parameterizations include diffusive mixing of water and ammonia vapors across the bottom of the model domain and a semi-gray radiative transfer scheme. The water and ammonia vapor abundances at the lower boundary are maintained to about three times the solar values via diffusive processes, where the diffusion coefficients come from those estimated by Visscher *et al*. [2010]. The semi-gray radiative transfer scheme considers solar heating and infrared (IR) cooling (*e.g.,* Schneider and Liu [2009]). The solar and IR optical depths are tuned to produce radiative fluxes similar to the Galileo probe measurements in a dry atmosphere between 0.4 to 20 bars (Sromovsky [1998]). Seasonal variation of solar insolation is ignored currently due to the very small obliquity of Jupiter, so is the diurnal variation since the rotation period (~ 10 hours) of Jupiter is much smaller than the typical dynamic time scale in the upper troposphere (~ 40 Earth days). The detailed formulations and configurations (e.g., tunable parameters) of the physics parameterizations will be presented in a paper dedicated to JupiterMPAS.

## 6.2 JupiterMPAS Model Setup

JupiterMPAS is run with a typical horizontal mesh spacing corresponding to about 0.5° (or equivalent to 620 km) and a vertical spacing of 5km. The horizontal spacing of a few hundred kilometers is required to resolve the synoptic-scale internal deformation radius that is typically a few thousand kilometers near equator on Jupiter [Vasavada and Showman, 2005; Lian and Showman, 2008]. There are 70 layers evenly spaced from 0.01 bars to 100 bars (~350 km in depth). The typical model time step is 480 seconds, which is about twice as large as those used in MITgcm-based Jupiter GCMs (Lian and Showman [2008, 2010]; Young *et al*. [2019a, 2019b]).

Initially there is no wind, and the reference temperature profile is similar to that in Lian and Showman [2008, 2010]. The water and ammonia vapor mass mixing ratios are initialized to be 85% of saturation values above their condensation levels calculated from the reference temperature profile, with deep abundances of 3× solar for both water and ammonia (*i.e.*, mass mixing ratios in regions deeper than water cloud base) (Li *et al*. [2017, 2020]).

## 6.3 JupiterMPAS Model Results

JupiterMPAS provides a similar or better match to observations compared with previous Jovian weather-layer dynamics models and has the most complete physics parameterizations. In this section we describe the model results in detail.

**6.3.1 Wind and Temperature**

JupiterMPAS successfully reproduces the equatorial superrotation and the banded, eastward-westward alternating zonal jet patterns (Fig. 16) with both water and ammonia hydrological cycles enabled. Figures 17a and 18b show the simulated zonal mean zonal winds and temperatures at the 0.5° mesh spacing needed to resolve the internal deformation radius of several thousand kilometers (*e.g.*, Lian and Showman [2008]). The equatorial superrotation has a typical speed of 170 m/s and extends from the tropopause (0.1 bars) down to the level near the water cloud base. Multiple banded jets develop in the weather layer and have barotropic components that extend all the way down to the bottom of the model domain at 100 bars.

As in Lian and Showman [2008, 2010], the equatorial superrotation is driven by baroclinic eddies that transfer energy to the zonal flows. Figure 18 shows the Eliassen-Palm (EP) fluxes (Fig. 18b, 18c), the divergence of which drives the equatorial superrotation (Fig. 18a). The EP fluxes at the equator are dominated by the eddy momentum flux component (Fig. 18b), therefore restricting eddy-mean flow interactions to be mostly in the 2D regime. The role of eddy energy in driving the equatorial superrotation can be further confirmed by spectral analysis of the eddy kinetic energy (EKE) via spherical harmonic decomposition of mass weighted and vertically averaged EKE between 0.1 and 3 bars. Figure 19 shows the zonal mean ($E_Z$) and residual ($E_R$) components of the power spectrum obtained from the spherical harmonic coefficients as a function of spherical wavenumber $n$ (Wieczorek and Meschede [2018]). The EKE at large horizontal scales (small wavenumbers) is dominated by the zonal mean flow as suggested by the ratio between $E_Z$ and $E_R$ for wavenumbers less than 10. For the zonal component $E_Z$, the slope of the power spectrum roughly follows the Kolmogorov form $n^{-5/3}$ for wavenumbers smaller than 20, and becomes $n^{-5}$ for wavenumbers larger than 40. For the residual component $E_R$, the slope is quite flat for small wavenumbers but turns to $n^{-5/3}$ when wavenumbers exceed 40. The $E_Z$ and $E_R$ spectra following an $n^{-5/3}$ depict the upscale flux of eddy energy to zonal mean flow and to larger scale eddies respectively. These results are consistent with the eddy kinetic energy spectrum of the Jovian cloud layer in Cassini images (Galperin et al. [2014]).

Globally (except for the equatorial region), the temperature in JupiterMPAS follows a dry adiabat in the deep atmosphere until it reaches the water cloud base, above which it becomes statically stable primarily due to the latent heating associated with the water moist convection (Fig. 20a). Between the water cloud top and ammonia cloud base, the temperature becomes nearly neutrally stable again. This behavior of the thermal profile, *i.e.*, a stable layer at a few bars, sandwiched between neutrally stable layers, is consistent with previous model results (*e.g.*, the cases with 2-3 times solar water abundance in Showman and Ingersoll 1998; Nakajima et al., 2000; Li and Chen, 2019). The stable layer below the water cloud base extends as deep as 8 bars, which acts as a barrier that prevents the ammonia rich air from entraining into the region above the water cloud base, except for the transport of ammonia via strong convection (high CAPE) such as thunderstorms (Showman and de Pater [2005]). In general, the model produces a pole-to-equator temperature contrast of a few Kelvins in the weather layer, which results from the combined effects of solar insolation and latent heating associated with moist convection. The magnitude of latitudinal temperature contrast near the tropopause [Fig. 20b] is comparable to that observed by Voyager IRIS and Cassini CIRS (e.g., Simon et al. [2006]).

### 6.3.2 Water and Ammonia

Water vapor in JupiterMPAS is mostly well-mixed from the bottom up until it reaches its condensation level near 7 bars, above which it is close to a saturation profile (Fig. 18c). There is a slight enrichment right below the cloud deck near the equator, as a result of re-evaporation of condensates such as rain and snow. Ammonia vapor exhibits depletion patterns like those in the Juno observations, although the magnitude of depletion is much weaker in JupiterMPAS and the depletion is mostly limited to the region above the water cloud base (Fig. 18d), whereas in reality it extends much deeper to 40-60 bars. Similar to the Juno observations, ammonia-rich air is present in a narrow latitudinal zone $\pm 5°$ at equator, and ammonia-poor air is located at mid-to-low latitudes.

### 7. Summary

We have converted a state-of-art atmospheric dynamical model, MPAS, a GCM using finite-volume methods that solves the fully-nonhydrostatic dynamic equations on unstructured grid, to simulate tropospheric and stratospheric dynamics in various planetary atmospheres. Benchmarks show that MPAS is computationally more efficient at fine horizontal grid resolutions, and highly scalable with large number of computer processors compared to WRF (a proxy for traditional latitude-longitude grid models). It is also numerically more stable than the MITgcm cube-sphere grid model at high resolution. Currently, Mars and Jupiter instantiations of planetMPAS have been developed with sophisticated physics suites and have been preliminarily tested again observations and prior modeling.

Before MPAS could be adopted for certain categories of planetary atmospheric application, it was necessary to consider the "pressure cooker" effect that results from the use of a height-based vertical coordinate, as opposed to the more widely used pressure-based vertical coordinate. The pressure cooker effect occurs since the air mass in a vertical column in a height-based model is bounded by a solid bottom and top, meaning that since density is therefore constant, the air pressure in the column would increase/decrease when temperature increases/decreases. This may be important for the Martian atmosphere (and many other atmosphere) where there are large temporal temperature variations. We examine the "pressure cooker" effect using an idealized column model as well as MarsMPAS that includes detailed physics processes. We find that the "pressure cooker" effect is negligible as long as the model top is sufficiently deep relative to the scale height, *e.g.*, the surface pressure error for Mars is order-of-magnitude of 0.01% if the model top is placed at 100 km above the surface. For Jupiter, this is of little concern since the atmospheric temperature changes by only a few Kelvin relative to the environment during moist convection events in sporadic regions. Nonetheless, both MarsMPAS and JupiterMPAS require sponge layers near the model top to minimize the wave reflection by the solid boundaries.

MarsMPAS shares most of its physics parameterizations with MarsWRF, including a K-distribution radiative transfer model with various radiatively active aerosols such as dust, water ice and $CO_2$ ice; surface and subsurface scheme that deal with Martian surface thermal properties modified by $CO_2$ ice; planetary boundary layer scheme that describes the convection as a diffusive process; radiatively active dust models that describe dust lifting and transport. Different from MarsWRF where the $CO_2$ cycle is parameterized as latent heat exchange between atmospheric column and surface, MarsMPAS treats $CO_2$ condensates in the air as an active tracer with prescribed mean particle sizes. An orographic gravity wave drag parameterization is also implemented in MarsMPAS, and has yet to be adapted to MarsWRF.

MarsMPAS is able to reproduce several key features in the Martian atmosphere compared to the observations and previous model results. These features include the seasonal variation of the $CO_2$ cycle characterized by the Viking lander pressure measurements, zonal mean temperature retrieved from MGS TES and MRO MCS, polar Argon enrichment and depletion measured by GRS on board the Mars Odyssey spacecraft, as well as some structural pattern of the density-scaled dust opacity derived from MRO MCS measurements. Compared to MarsWRF, MarsMPAS produces better polar warming and far better Argon enrichment due to improved physical and numerical representations of the polar circulations.

JupiterMPAS is developed following successful modeling of Jovian atmospheres using the MITgcm. The physics parameterizations implemented in JupiterMPAS represent a significant improvement from the MITgcm version of the Jupiter model that used idealized schemes such as Newtonian relaxation. The parameterizations in JupiterMPAS include dry convective mixing of temperature, momentum and tracers in the sub-cloud layers, a mass-flux based moist convection scheme for water, a Kessler-type cloud microphysics model for water, a Betts-Miller type moist convection scheme for ammonia, an improved method of terminal velocity calculation and a gray radiative transfer model that represents solar heating and infrared cooling in the atmosphere.

JupiterMPAS is able to reproduce large-scale dynamical features at the visible cloud top in the Jovian atmosphere, including the equatorial superrotation with speed comparable to the cloud tracking results, banded zonal jet patterns at various latitudes, and the few Kelvin pole-to-equator temperature contrast. Like previous simulations with the MITgcm, the zonal winds in JupiterMPAS extend all the way down to the bottom of the simulation domain. The jet driving mechanisms are similar to the observations and previous model results, which show that the zonal jets including the equatorial superrotation are primarily driven by inverse cascade of turbulent energy.

The overall development of planetMPAS has been a success in terms of its ability to reproduce observations and previous model results. We are actively expanding the capability of planetMPAS, such as its application to slow-rotating planets Titan and Venus. Further effort on the model development will include testing the active water cycle for Mars, increasing the model top to cover the lower thermosphere for Mars and Titan, and potentially a coupled deep-shallow atmospheric model for Jupiter. The latter two components will utilize the recently improved MPAS dynamical core that properly treats the vertical Coriolis forces and curvature terms associated with the vertical momentum equation.

**Acknowledgement**: The development of MarsMPAS is supported by NASA grant 80NSSC20K0468 and MFRP-NNX16AK98G. The development of JupiterMPAS is supported by NASA grant 80NSSC18K0088.

Betts, A. K. and Miller, M. J., 1986. "A new convective adjustment scheme, Part II: Single column tests using GATE wave, BOMEX, ATEX and arctic air-mass data sets", *Quarterly Journal of the Royal Meteorological Society*, vol. 112, no. 473, pp. 693–709. doi:10.1002/qj.49711247308.

Brecht, A.S., Kahre, M.A., Kling, A.L.M., Wilson, R.J., & Hollingsworth, J.L. (2019). Examining seasonal trends of the martian polar warming with the NASA Ames Mars Global Climate Model [Abstract]. EPSC-DPS Joint Meeting 2019. Geneva, Switzerland, 15-20 September.

Bolton, S.J., Adriani, A., Adumitroaie, V., Allison, M., Anderson, et al., 2017. Jupiter's interior and deep atmosphere: The initial pole-to-pole passes with the Juno spacecraft. *Science* 356, 821–825.

Bougeault, P. and Lacarrère, P., 1989. "Parameterization of orographic induced turbulence in a mesobeta scale model", *Monthly Weather Review*, vol. 117, pp. 1872–1890. doi:10.1175/1520-0493(1989)117<1872:POOITI>2.0.CO;2.

Bougher, S. W., et al. (2017), The structure and variability of Mars dayside thermosphere from MAVEN NGIMS and IUVS measurements: Seasonal and solar activity trends in scale heights and temperatures, *J. Geophys. Res. Space Physics*, 122

Chen-Chen, H. S. Perez-Hoyos, and A. Sanchez-Lavega (2019). Dust Particle size and optical depth on Mars Retrieved by the MSL navigation cameras. *Icarus* 319, 43-57.63

Clancy, R. T., Wolff, M. J., Christensen, P. R. (2003). Mars aerosol studies with the MGS TES emission phase function observations: Optical depths, particle sizes, and ice cloud types versus latitude and solar longitude. *Journal of Geophysical Research* 108, doi: 10.1029/2003JE002058

Colaïtis, A., Spiga, A., Hourdin, F., Rio, C., Forget, F., & Millour, E. (2013). A thermal plume model for the Martian convective boundary layer. Journal of Geophysical Research: Planets, 118(7), 1468-1487.

Colaprete, A., Barnes, J. R., Haberle, R. M., and Montmessin, F., 2008. "$CO_2$ clouds, CAPE and convection on Mars: Observations and general circulation modeling", *Planetary and Space Science*, vol. 56, no. 2, pp. 150–180. doi:10.1016/j.pss.2007.08.010.

Collins, M., S.R. Lewis, and P.L. Read, 1997. Gravity wave drag in a global circulation model of the Martian atmosphere: Parameterisation and validation, *Adv. Space Res.*,19(8), 1245-1254

Creasey, J. E., J. M. Forbes, and D. P. Hinson (2006a), Global and seasonal distribution of gravity wave activity in Mars' lower atmosphere derived from MGS radio occultation data, *Geophys. Res. Lett.*, 33, L01803

Creasey, J. E., J. M. Forbes, and G. M. Keating (2006b), Density variability at scales typical of gravity waves observed in Mars' thermosphere by the MGS accelerometer, *Geophys. Res. Lett.*, 33, L22814

Durran, D. R. (1999). Numerical Methods for Wave Equations in Geophysical Fluid Dynamics Springer. New York.
64

Judt, F., 2020. Atmospheric predictability of the tropics, middle latitudes, and polar regions explored through global storm-resolving simulations, *J. Atmos. Sci.*, 77, 257–276, https://doi.org/10.1175/JASD-19-0116.1.

Kaspi, Y., Galanti, E., Hubbard, W. *et al.*, 2018. Jupiter's atmospheric jet streams extend thousands of kilometres deep. *Nature* 555, 223–226.

Kaspi, Y., Flierl, G.R., Showman, A.P., 2009. The deep wind structure of the giant planets: Results from an anelastic general circulation model. *Icarus*, 202, 525-542.

Kessler, E., 1969. On the Distribution and Continuity of Water Substance in Atmospheric Calculations. *Meteorological Monographs* 32, 84.

Khvorostyanov, V.I., Curry, J.A., 2002. Terminal Velocities of Droplets and Crystals: Power Laws with Continuous Parameters over the Size Spectrum. *J. Atmos. Sci*. 59, 1872-1884.

Kim, Y.-J. and Arakawa, A., 1995. "Improvement of Orographic Gravity Wave Parameterization Using a Mesoscale Gravity Wave Model.", *Journal of Atmospheric Sciences*, vol. 52, no. 11, pp. 1875–1902. doi:10.1175/1520-0469(1995)052<1875:IOOGWP>2.0.CO;2.

Kim, Y.-J. and Doyle, J. D., 2005. "Extension of an orographic-drag parametrization scheme to incorporate orographic anisotropy and flow blocking", *Quarterly Journal of the Royal Meteorological Society*, vol. 131, no. 609, pp. 1893–1921. doi:10.1256/qj.04.160.
68

Ringler, T., J. Thuburn, J. B. Klemp, and W. C. Skamarock, 2010: A unified approach to energy conservation and potential vorticity dynamics for arbitrarily-structured C-grids. *J. Comput. Phys.*, 229, 3065–3090, https://doi.org/10.1016/j.jcp.2009.12.007.

Schneider, T., Liu, J., 2009. Formation of jets and equatorial superrotation on Jupiter. *J. Atmos. Sci.* 66, 579–601

Schneider, T., S. D. B. Graves, E. L. Schaller, and M. E. Brown, 2012: Polar methane accumulation and rainstorms on Titan from simulations of the methane cycle. *Nature*, 481, 58–61

Schoeberl, M. R., Strobel, D. F., and Apruzese, J. P., 1983. "A numerical model of gravity wave breaking and stress in the mesosphere", *Journal of Geophysical Research*, vol. 88, no. C9, pp. 5249–5259. doi:10.1029/JC088iC09p05249.

Shaposhnikov, D. S., Medvedev, A. S., Rodin, A. V., and Hartogh, P., 2019. "Seasonal Water "Pump" in the Atmosphere of Mars: Vertical Transport to the Thermosphere", *Geophysical Research Letters*, vol. 46, no. 8, pp. 4161–4169. doi:10.1029/2019GL082839.

Showman, A. P., Ingersoll, A. P., 1998. Interpretation of Galileo Probe Data and Implications for Jupiter's Dry Downdrafts. *Icarus*, 132, 205–220

Showman, A. P., and I. de Pater (2005), Dynamical implications of Jupiter's tropospheric ammonia abundance, *Icarus*, 174(1), 192–204, doi:10.1016/j.icarus.2004.10.004.

Showman, A. P., et al., 2009. "Atmospheric Circulation of Hot Jupiters: Coupled Radiative-
75

New Horizons encounter", *Icarus*, vol. 254, pp. 306–323. doi:10.1016/j.icarus.2015.03.034.

Toigo, A. D., Waugh, D. W., Guzewich, S. D., 2017. What causes Mars' annular polar vortices? *Geophys. Res. Lett*., 44, 71-78

Tyler, D., Barnes, J. R., and Haberle, R. M., 2002. "Simulation of surface meteorology at the Pathfinder and VL1 sites using a Mars mesoscale model", *Journal of Geophysical Research (Planets)*, vol. 107, no. E4. doi:10.1029/2001JE001618.

Vasavada, A.R., Showman, A.P., 2005. Jovian atmospheric dynamics: An update after Galileo and Cassini. *Rep. Prog. Phys*. 68, 1935–1996.

Visscher, C., Moses, J. I., Saslow, S. A., Oct 2010. The deep water abun- dance on Jupiter: New constraints from thermochemical kinetics and diffusion modeling. *Icarus* 209 (2), 602–615.

Wieczorek, M. A., & Meschede, M. (2018). SHTools: Tools for working with spherical harmonics. *Geochemistry, Geophysics, Geosystems*, 19, 2574–2592.

Wicker, L. J., and W. C. Skamarock, 2002: Time-splitting methods for elastic models using forward time schemes. *Mon. Wea. Rev*., 130, 2088–2097.

Wilson, R. J., & Hamilton, K. (1996). Comprehensive Model Simulation of Thermal Tides in the Martian Atmosphere, *Journal of Atmospheric Sciences*, 53(9), 1290-1326.
79

**Table 1:**

| | MPAS | | | | | WRF | | | | | MITgcm | |
|---|---|---|---|---|---|---|---|---|---|---|---|---|
| Horizontal grid points | 2,562 | 10,242 | 40,962 | 163,842 | 655,362 | 4,050 | 16200 | 64,800 | 259,200 | 1,036,800 | 2592 | 10368 |
| Equivalent horizontal resolutions (degree) | 4 | 2 | 1 | 0.5 (0.75-0.2) | 0.25 | 4 | 2 | 1 | 0.5 | 0.25 | 4 | 2 |
| Number of processors | 8 | 32 | 128 | 512 | 2048 | 8 | 32 | 128 | 512 | 2048 | 8 | 32 |
| Time step (seconds) | 60 | 60 | 60 | 30 (15) | 15 | 60 | 60 | 60 | 20* | 2* | 60 | 60 |
| Total Wall clock used per day (minutes) | 6.93 | 7.07 | 7.13 | 11 (17.60) | 20.33 | 5.74 | 6.34 | 7.25 | 18.28 | 243.29 | 4.85 | 5.83 |
| Total Wall clock per time step (seconds) | 0.289 | 0.295 | 0.297 | 0.229 (0.183) | 0.212 | 0.229 | 0.253 | 0.288 | 0.250 | 0.334 | 0.202 | 0.243 |
| Dynamics Wall clock per time step (seconds) | 0.111 | 0.117 | 0.119 | 0.129 (0.124) | 0.152 | 0.089 | 0.109 | 0.144 | 0.200 | 0.330 | 0.07 | 0.072 |
| Atmospheric layers | 42 | | | | | | | | | | | |

| Subsurface layers | 12 |
|---|---|
| Physics | KDM RT, YSU PBL, Subsurface model, CO2 cycle |
| Number of Tracers | 2 |
| System configuration | NASA Pleiades Haswell Nodes<br><br>Processors: Intel Xeon E5-2680v3 (2.5GHz)<br><br>Processors per node: 2x12cores |

**Table 1.** Computational performance of MPAS at various resolutions and as the number of parallel processors is increased commensurate with the resolution such that the number of grid points per processor remains nearly constant (at roughly 320 grid points per processor). Note that all except one MPAS simulation uses meshes that have approximately constant horizontal spatial resolution. For the 163,842 grid point case, two different meshes are shown: the constant 0.5 degree (30 km) resolution case (without parentheses) and a zoomed mesh with resolution varying from 0.75-0.2 degrees (45-12 km) resolution (values shown in parentheses). Reference simulations at similar equatorial horizontal resolution are shown for the WRF and MITgcm models using the same Mars "physics" packages. The WRF simulations marked with an asterisk (*) are probably overoptimistic for the timestep for numerical stability for simulations as only 1 day was simulated from spin-up; a much shorter timestep is likely required for a viable spun-up case (a prior, stable, spun-up 0.2 degree global WRF simulation was found to require a 0.25 s timestep). Note also that while only two tracers were active in all simulations, but MPAS carries and operates upon a default of 14

tracers and as such the number of calculations done by MPAS is greater than WRF or MITgcm; MPAS uses a monotonic flux limiter, WRF does not.

**Table 2:**

|  | MPAS (2-degree) | MPAS (2-degree) | MPAS (2-degree) |
|---|---|---|---|
| Number of processors | 16 | 32 | 64 |
| Time step(seconds) | 120 | | |
| KDM RT update frequency (minutes) | 30 | | |
| Total Wall clock used per day (minutes) | 6.75 | 3.14 | 1.49 |
| Total Wall clock used per time step (seconds) | 0.562 | 0.262 | 0.124 |
| Halo exchange Wall clock used per time step (seconds) | 0.130 | 0.063 | 0.029 |
| Dynamics Wall clock per time step (seconds) | 0.241 | 0.117 | 0.061 |
| Dynamics Wall clock per timestep * number of processors (proc-seconds) | 3.856 | 3.744 | 3.904 |

**Table 2.** The performance of a fixed-resolution MPAS simulation as the number of processors is increased.

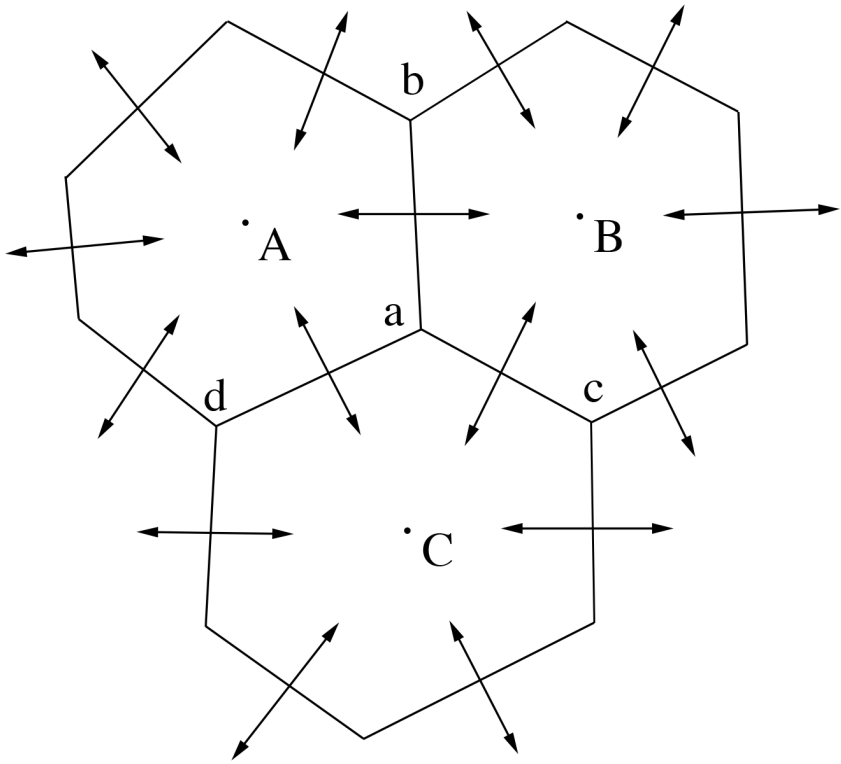

**Figure 1.** The unstructured Voronoi mesh (or centroidal Voronoi tessellation, CVT), with cell centers (A, B, C) and vertices (a, b, c, d). The arrowed lines connecting the cell centers represent the velocity vectors, with the velocity values calculated at the points where the velocity vectors intersect the cell edges (ab, ac, ad) at right angles. MPAS uses a finite volume numerics representing prognostic properties within each cell and fluxes across the cell walls.

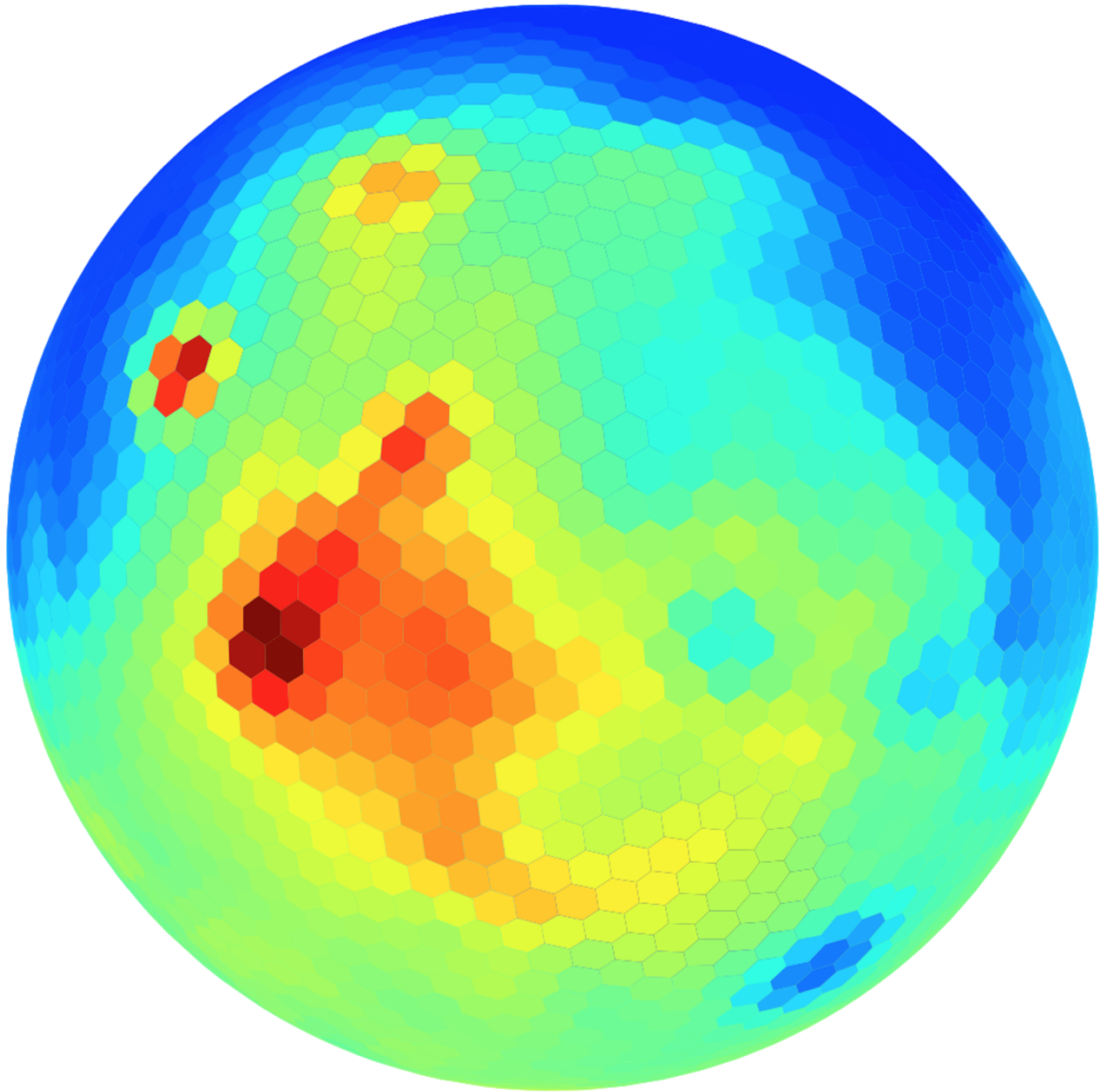

**Figure 2.** Mars Orbiter Laser Altimeter (MOLA) topography projected onto the lowest resolution spherical centroidal Voronoi tessellation (SCVT) mesh used in this paper (a 2,562 cell mesh that approximates 4 degree latitude and longitude resolution near the equator). The very low-resolution map is shown so that the individual polygonal cells, shown in schematic form in Figure 1), can be identified in the global projection. The nearly uniform global sampling by the mesh should also be noted.

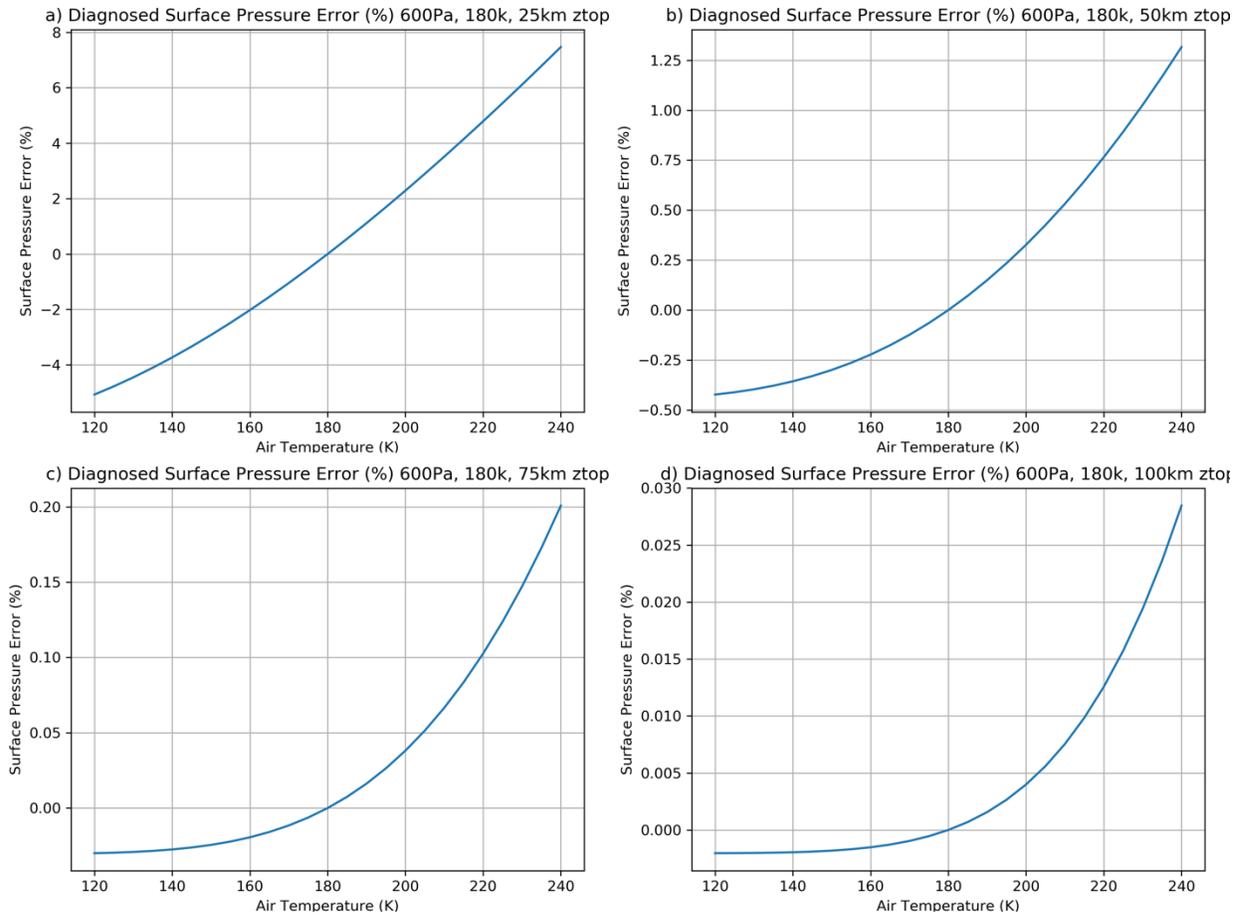

**Figure 3.** Error in diagnosed surface pressure for an idealized, laterally uniform model with a rigid lid and a hydrostatic vertical mass distribution when the isothermal atmospheric temperature is changed from an initial reference of 180K. See text for a more complete description of the idealized model. Model uses Mars gravity, Mars gas constants, and an initial surface pressure of 600 Pa. Cases are shown with the rigid lid at (a) 25km, (b) 50km, (c) 75km, and (d) 100km. Note the substantial changes in the y-axis values between plots.

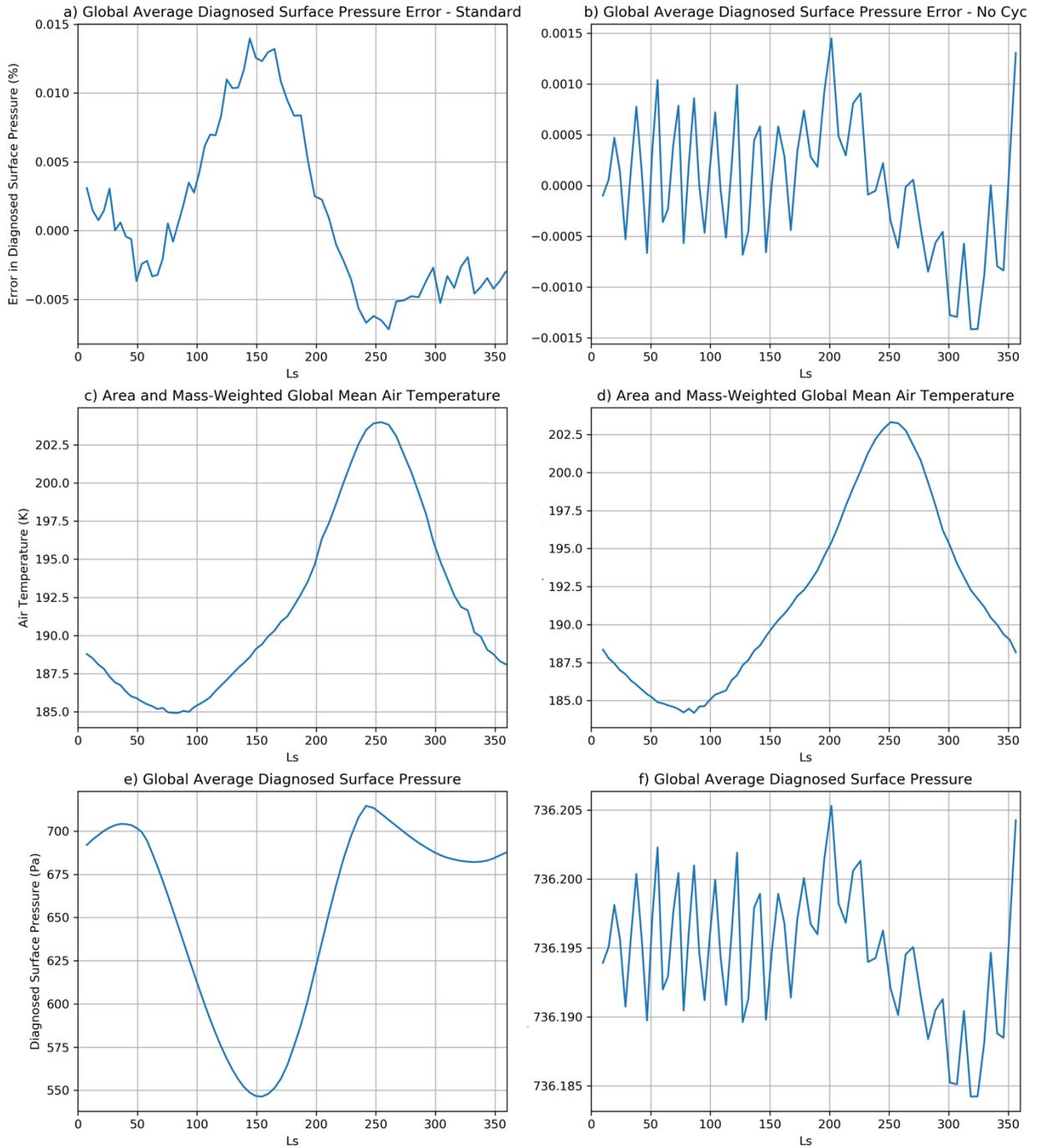

**Figure 4.** MarsMPAS simulations for cases (left hand column) with and (right hand column) without an active $CO_2$ cycle. All output are shown as a function of the seasonal date, $L_s$, measured in degrees, and with northern spring equinox at $L_s=0°$, northern summer equinox at $L_s=90°$, *etc*. Panels (a) and (b) show the error in global average diagnosed surface pressure relative to the known

total initial mass of $CO_2$. Panels (c) and (d) shown the global area-weighted and vertical mass-weighted average air temperature. Panels (e) and (f) show the global average diagnosed surface pressure.

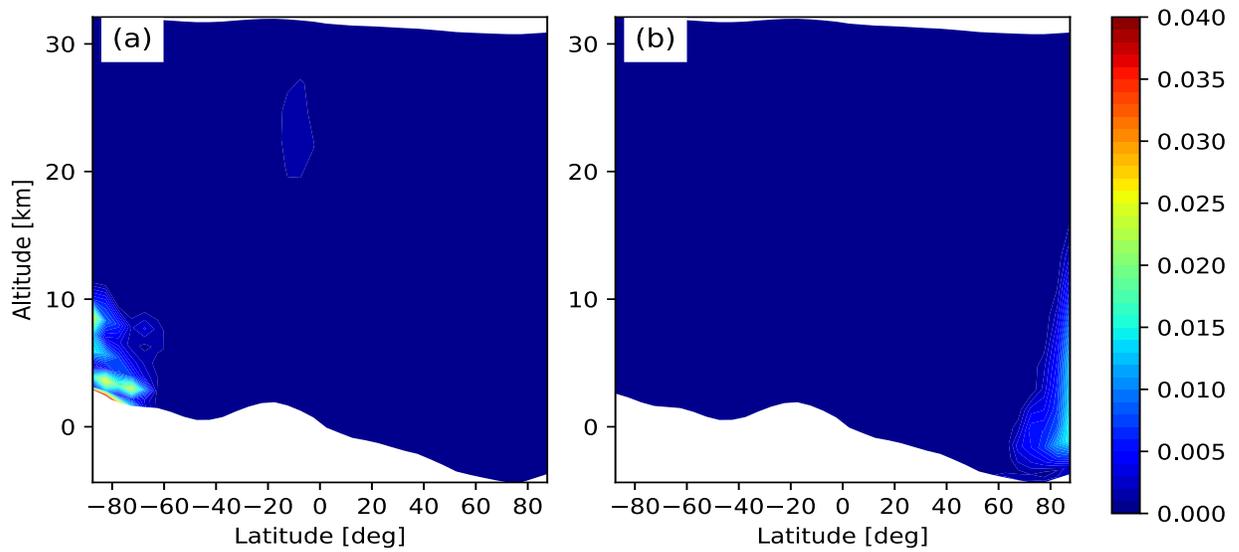

Figure 5. Zonal mean mass mixing ratio of $CO_2$ ice in air: (a) $L_s = 90°$; (b) $L_s = 270°$. The unit is [kg/kg].

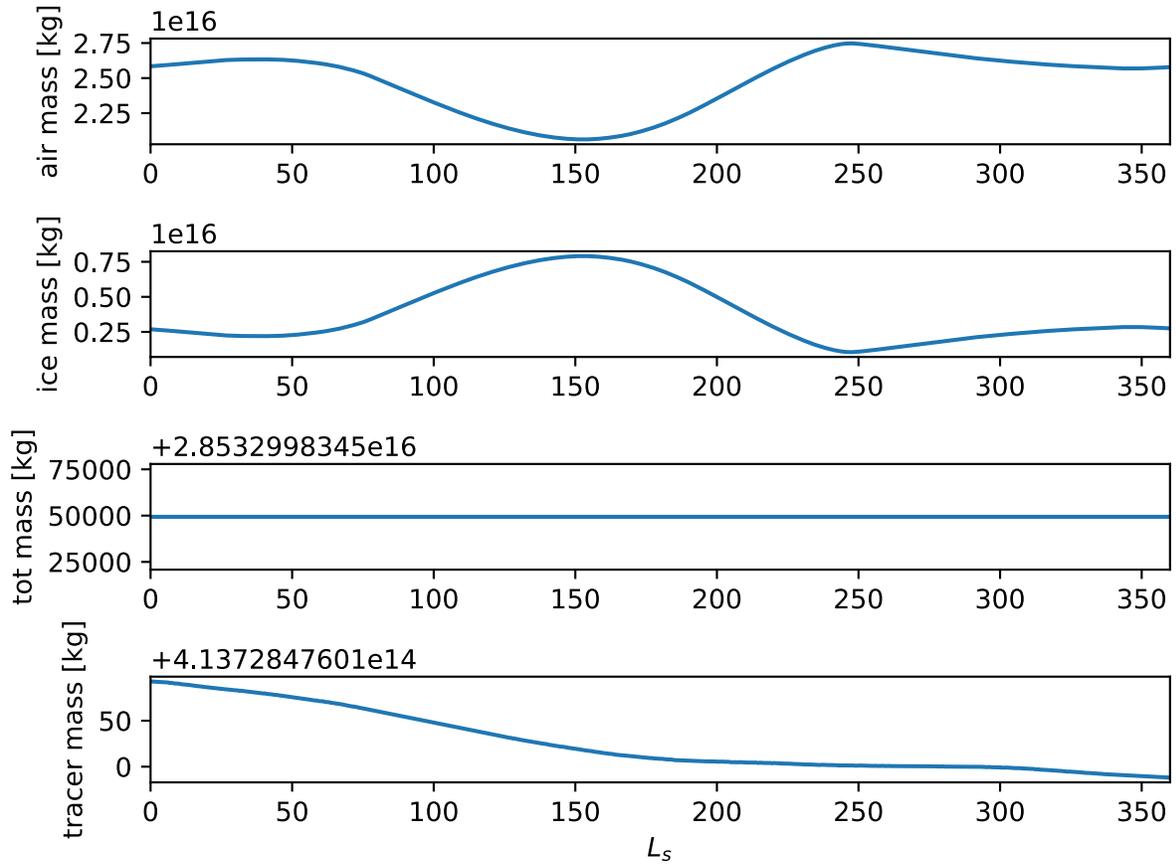

Figure 6. Mass conservation in MarsMPAS. From top down: $CO_2$ air mass; $CO_2$ ice mass (surface ice and ice particles in air); total $CO_2$ mass (air+ice) and total Argon mass.

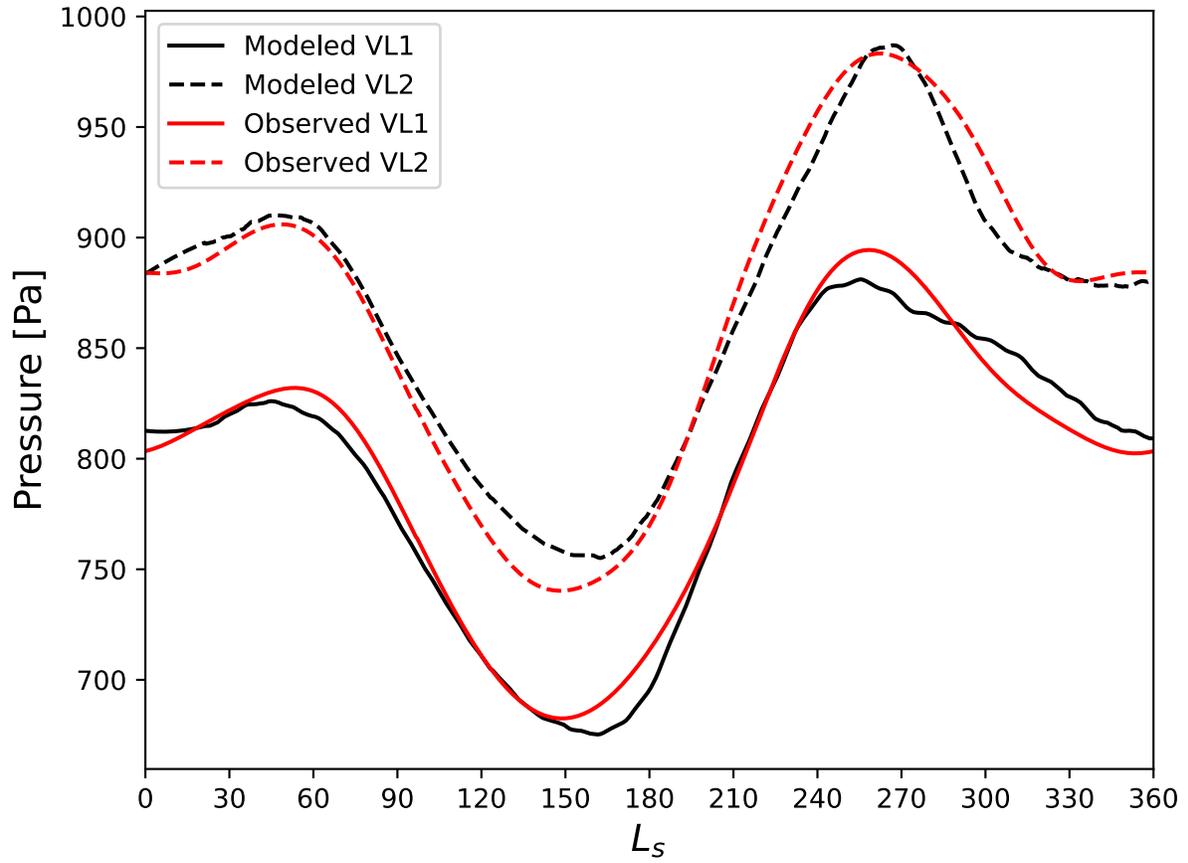

Figure 7. Modeled $CO_2$ cycles (black lines) compared to the Viking Lander measurements (red lines). The modeled and the measured $CO_2$ cycles are both fitted over a Martian year.

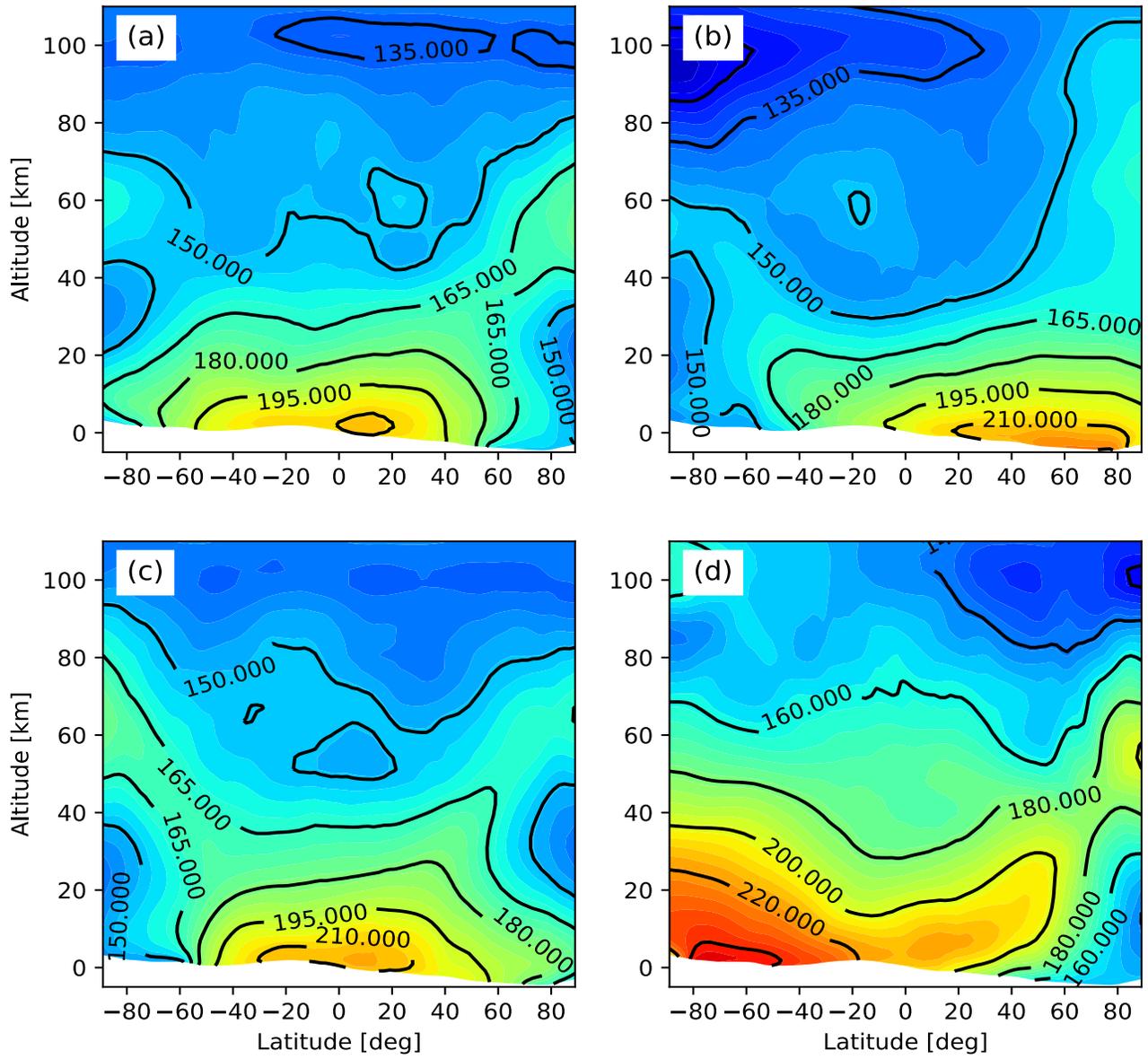

Figure 8. Zonal mean temperature at $L_s = 0°$ (a), $L_s = 90°$ (b), $L_s = 180°$ (c) and $L_s = 270°$ (d).

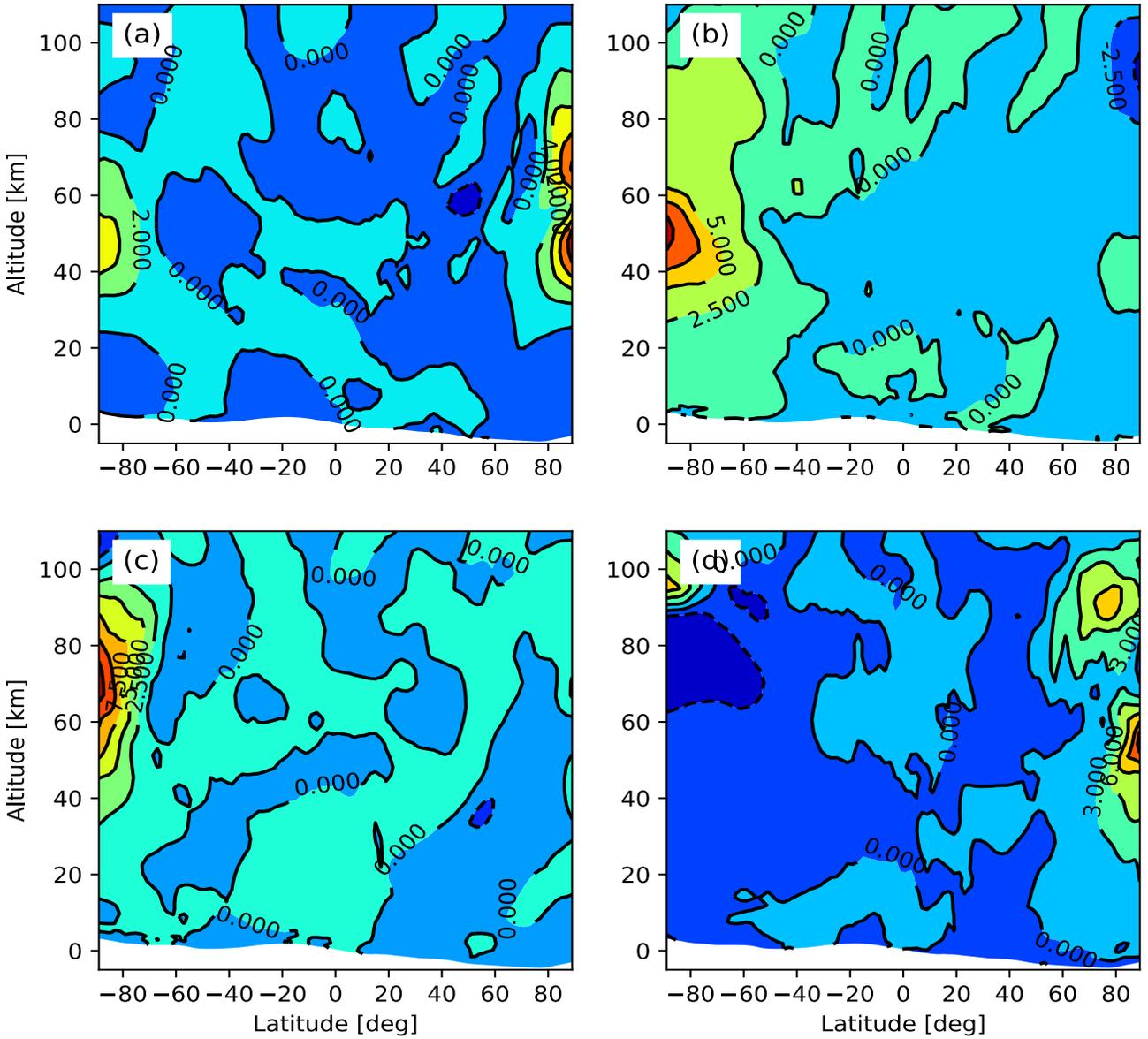

Figure 9. Similar to Figure 8 but the figure shows the temperature difference between the cases with and without GWDO. It can be seen that the presence of GWDO leads to a few Kelvin polar warming in both hemispheres.

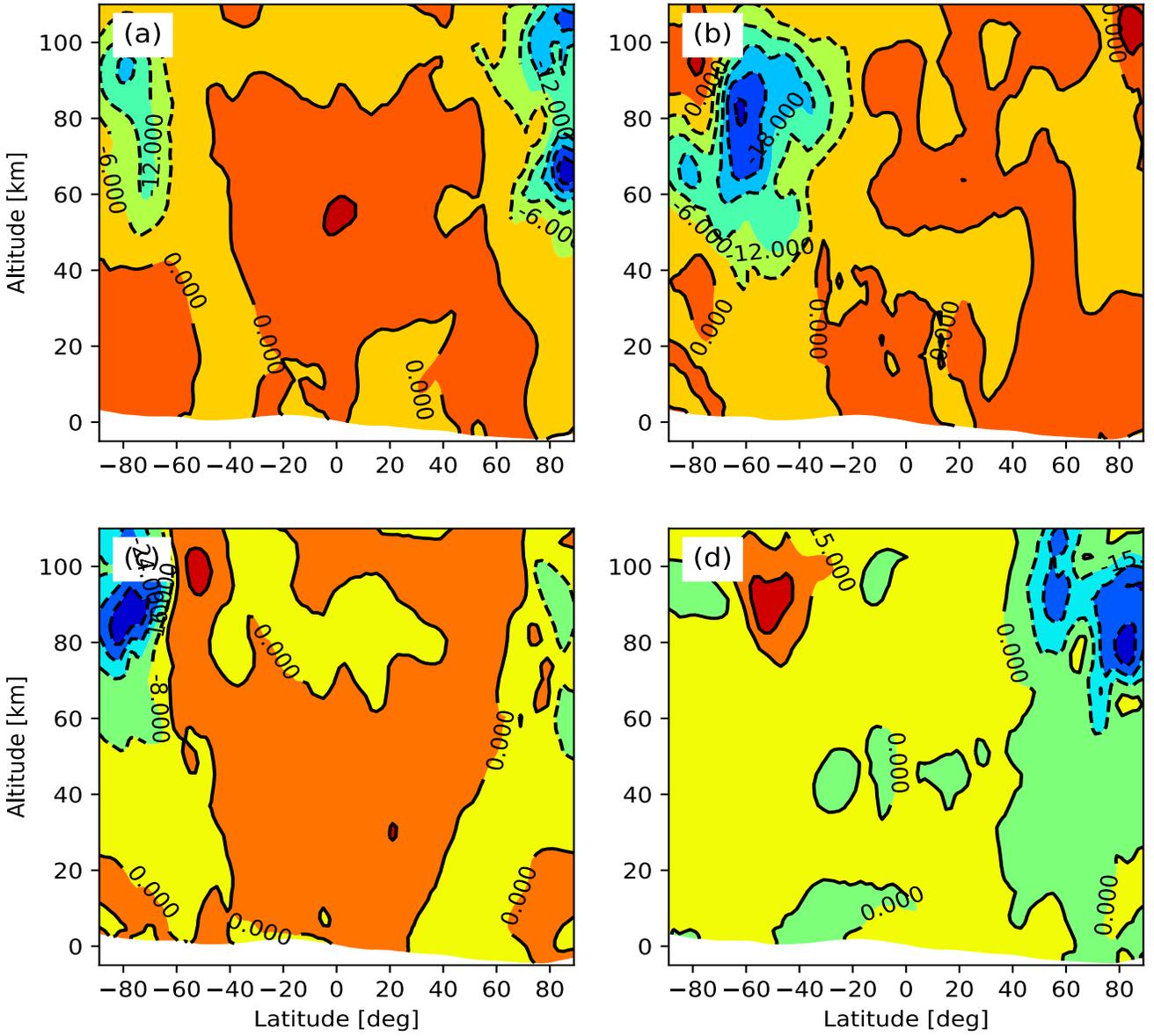

Figure 10. Similar to Figure 9 but the figure shows the zonal wind difference between the cases with and without GWDO.

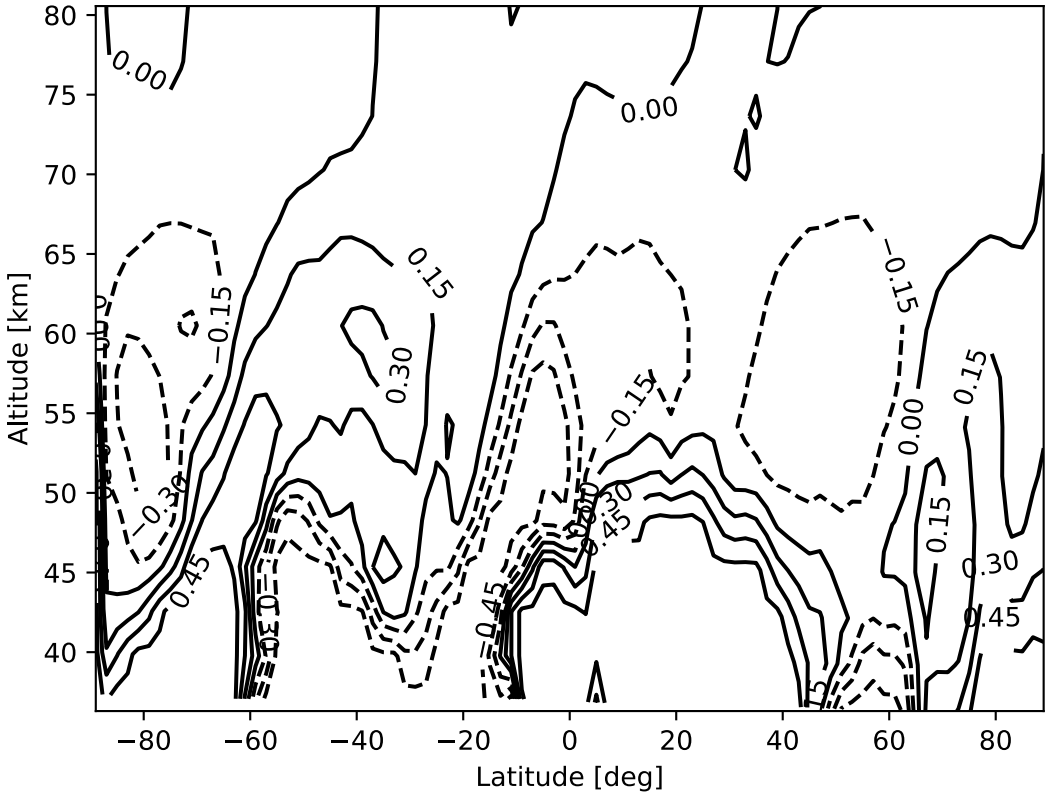

Figure 11. The difference of residual mean circulations between the cases with and without GWDO at $L_s = 90°$. The contour lines show the streamfunctions with unit of kg $m^{-1}s^{-1}$. The dashed lines mean counter-clockwise rotation, and the solid lines mean clockwise rotation.

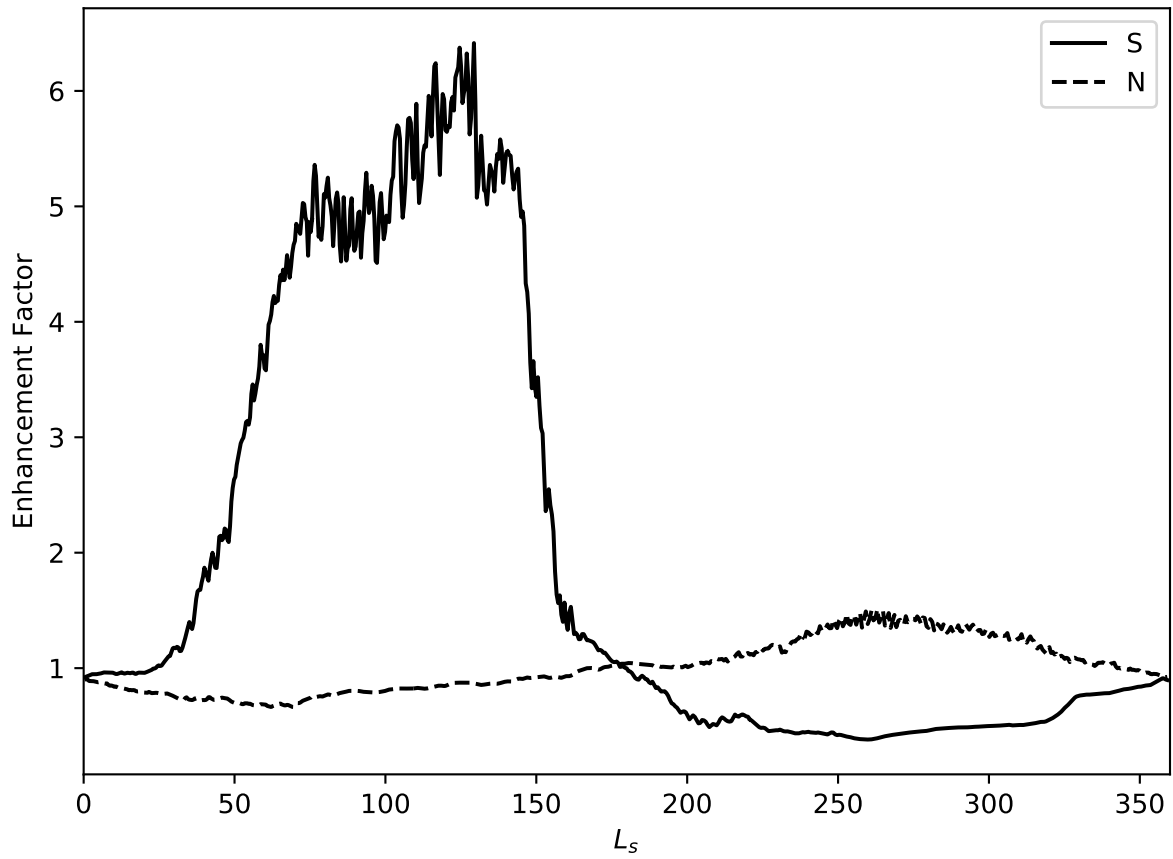

Figure 12. Argon enhancement factor in the south polar region (solid line) and north polar region (dashed line) over a Martian year.

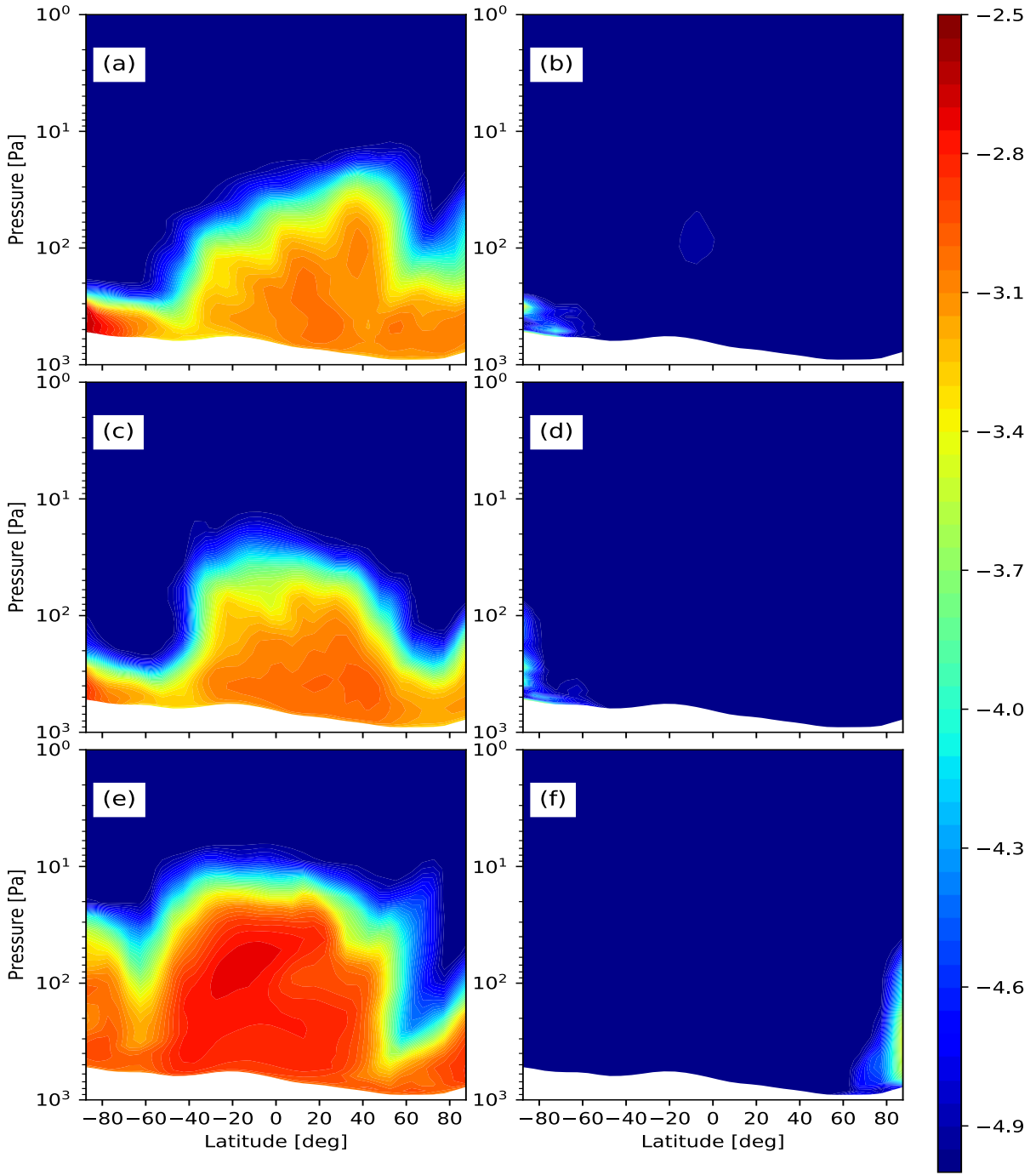

Figure 13. $Log_{10}$ zonal average density-scaled dust opacity (a, c, e) and $CO_2$ ice opacity (b, d, f) at $L_s = 90°$ (a, b), $L_s = 150°$ (c, d) and $L_s = 300°$ (e, f).

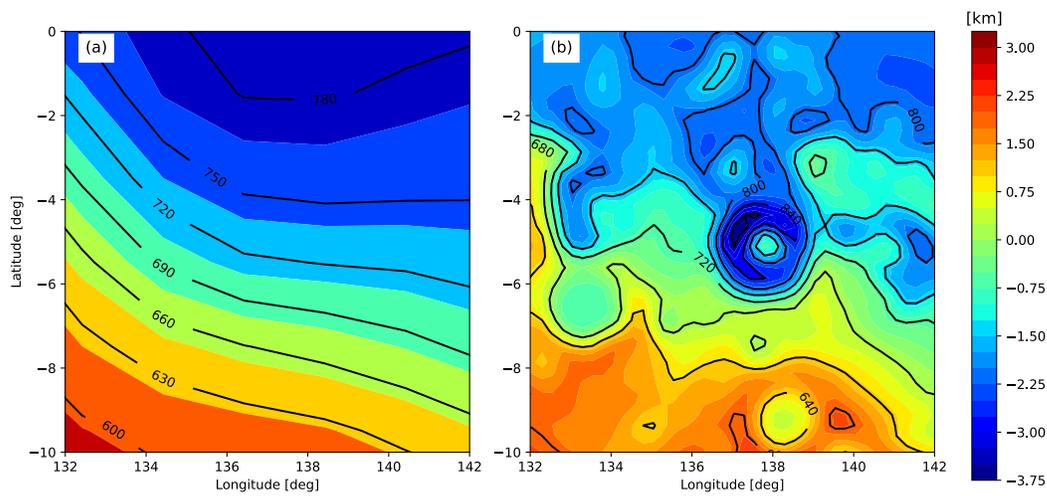

Figure 14. Surface pressure simulated by uniform 2° resolution (a) and variable resolution 0.75° - 0.25° (b) in the vicinity of Gale Crater near $L_s = 60°$. The colormaps are the terrain elevation in unit of kilometers, and the black curves are pressure with steps of 30 Pa in (a) and 40 Pa in (b) between adjacent contour lines.

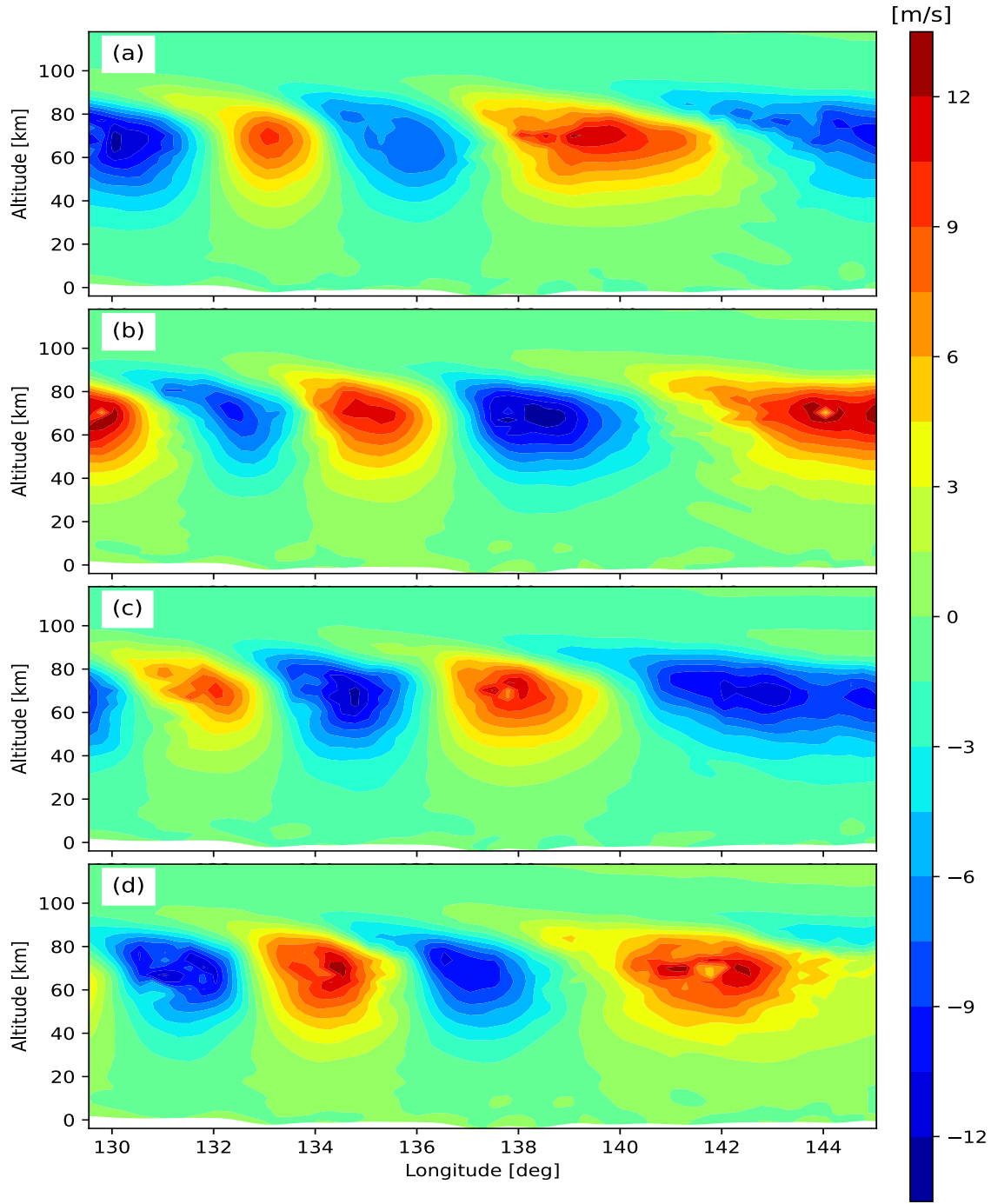

Figure 15. Longitudinal cross sections of vertical velocity at latitude $\phi = -4.49°$ (centered at Gale Crater) near $L_s = 60°$. The plots (a), (b), (c) and (d) are generated by an output frequency of one hour.

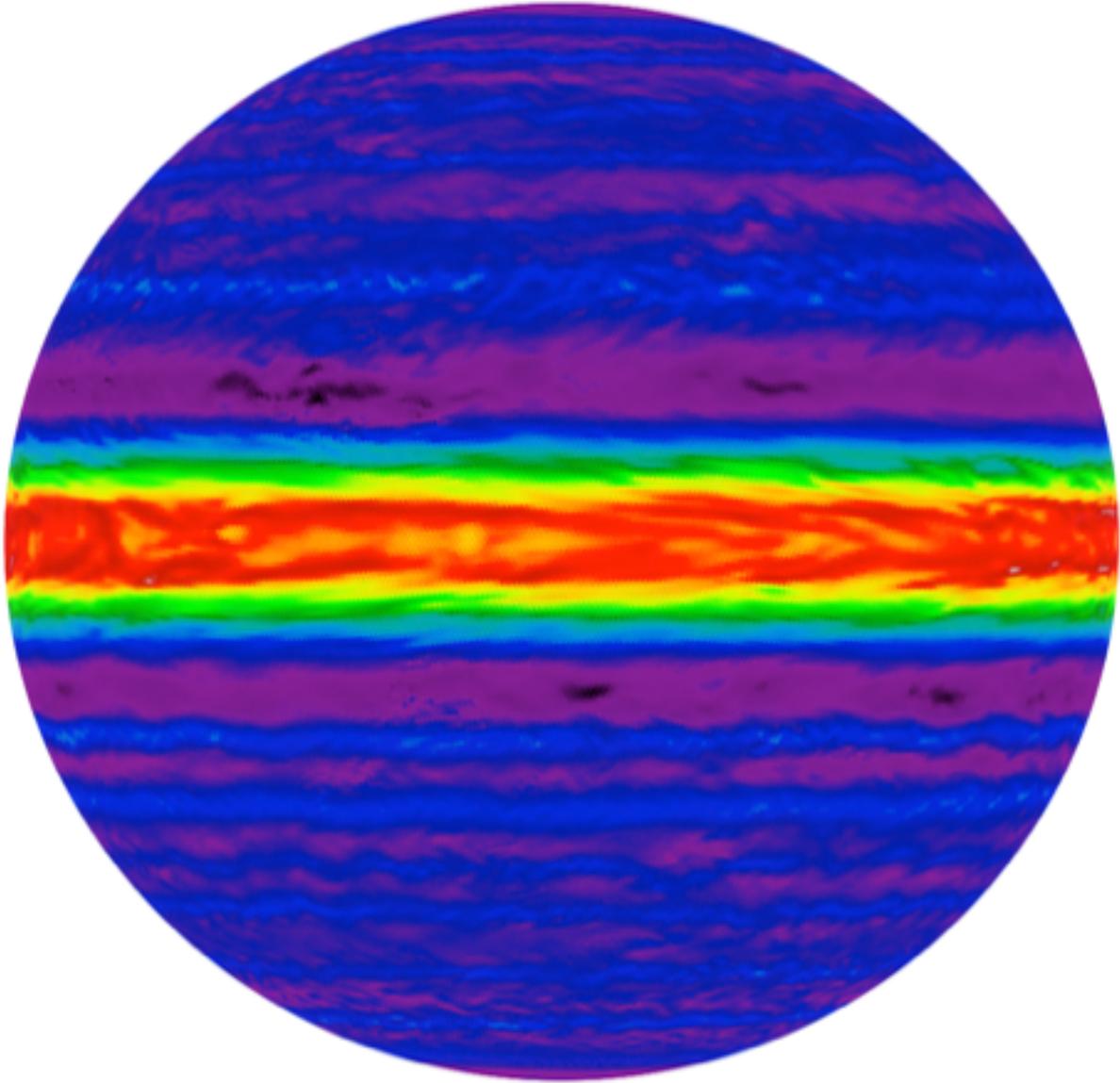

Figure 16. Zonal wind at 1 bar simulated in JupiterMPAS. The speed of the equatorial superrotation is ~ 170 m/s at equator (red color). The wind speed outside of the ±5° equatorial band is about 20m/s (purple color for eastward flow and blue for westward flow).

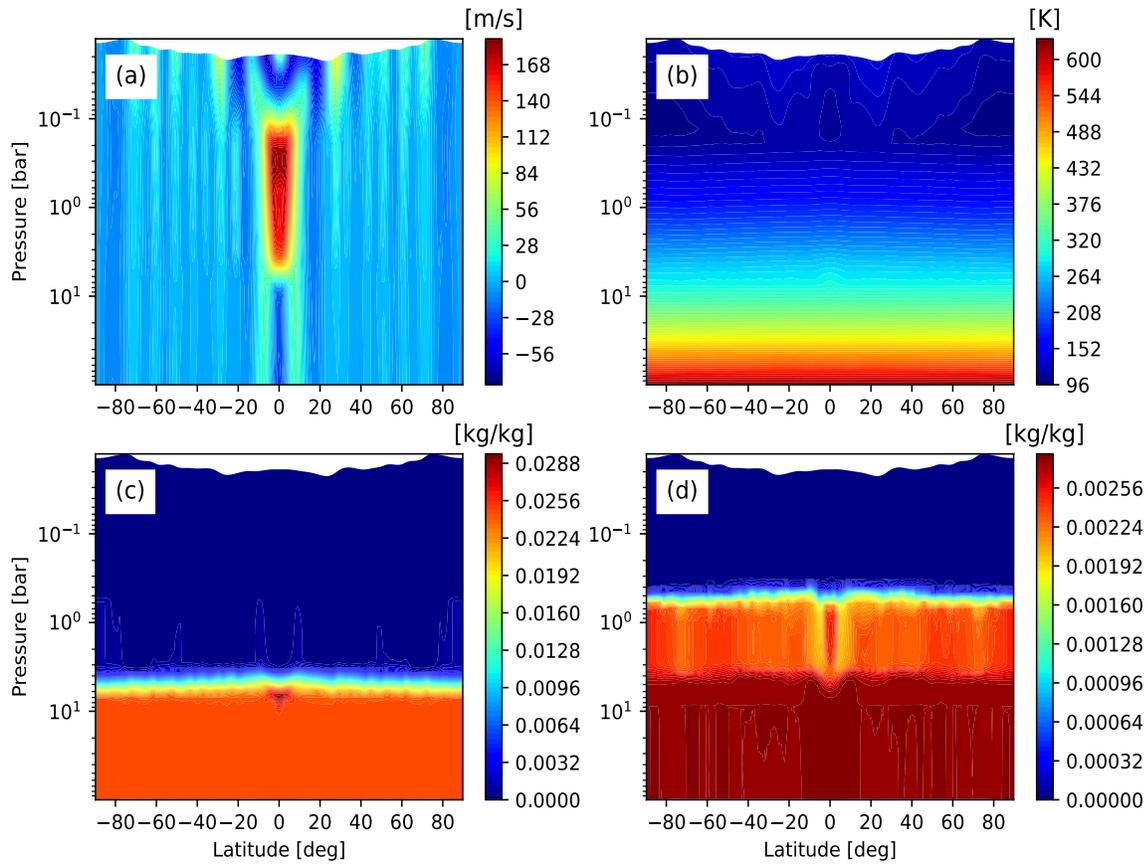

Figure 17. Zonal mean zonal wind (a), temperature (b), water vapor mixing ratio (c) and ammonia vapor mixing ratio (d).

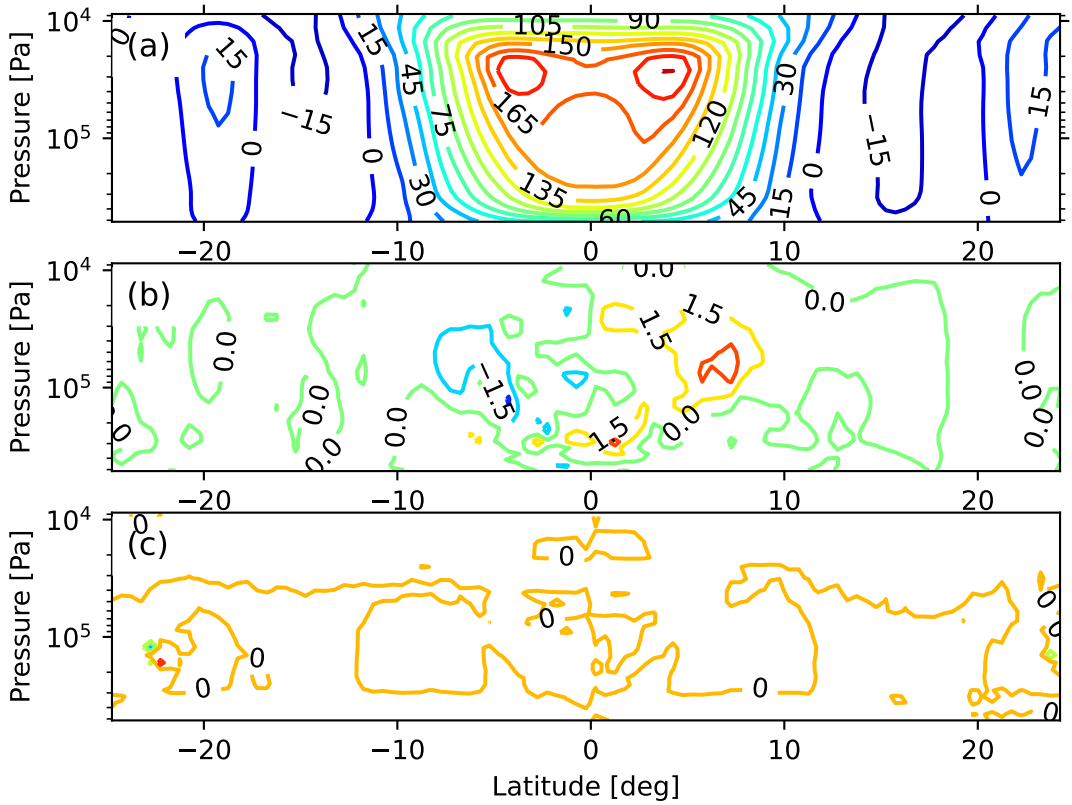

Figure 18. Zonal mean zonal wind (a) and EP Fluxes including eddy momentum flux (b) and eddy heat flux (c).

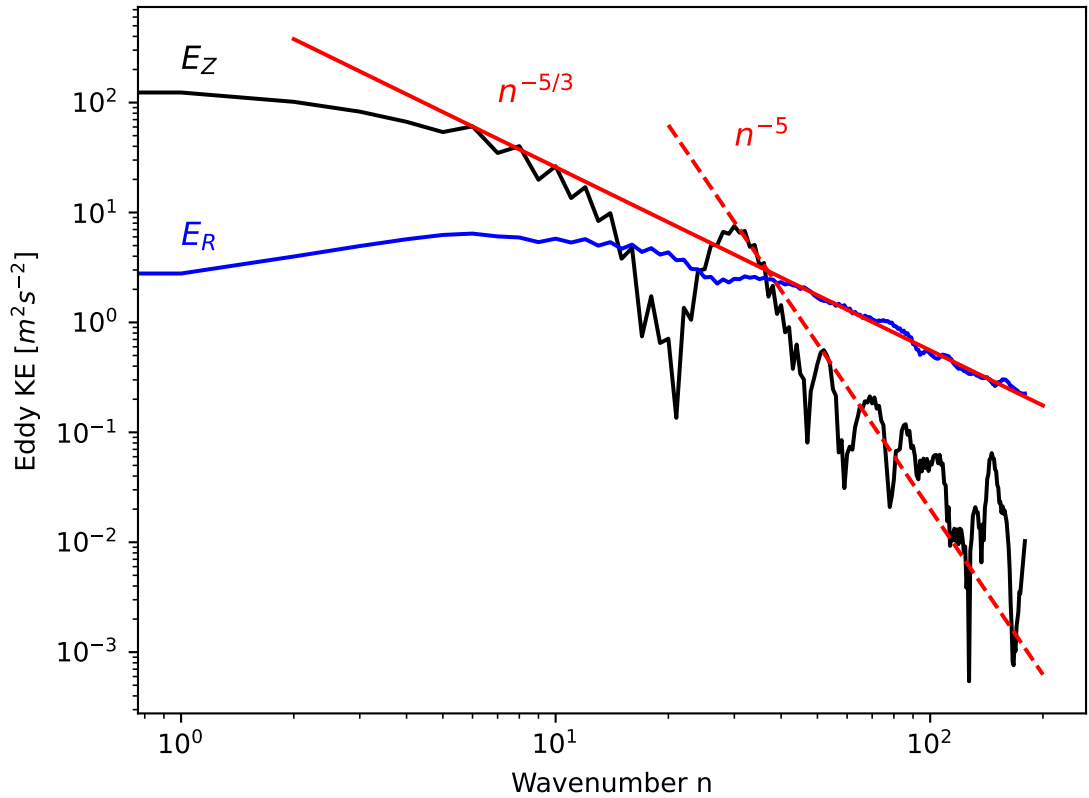

Figure 19. Power spectrum of zonal and residual eddy kinetic energy (solid black and blue lines) as a function of spherical wavenumber. The eddy kinetic energy is vertically averaged and mass weighted between 0.1 bars and 3 bars. The red solid and dashed lines show the Kolmogorov form $n^{-5/3}$ and $n^{-5}$ respectively, where $n$ is the wavenumber.

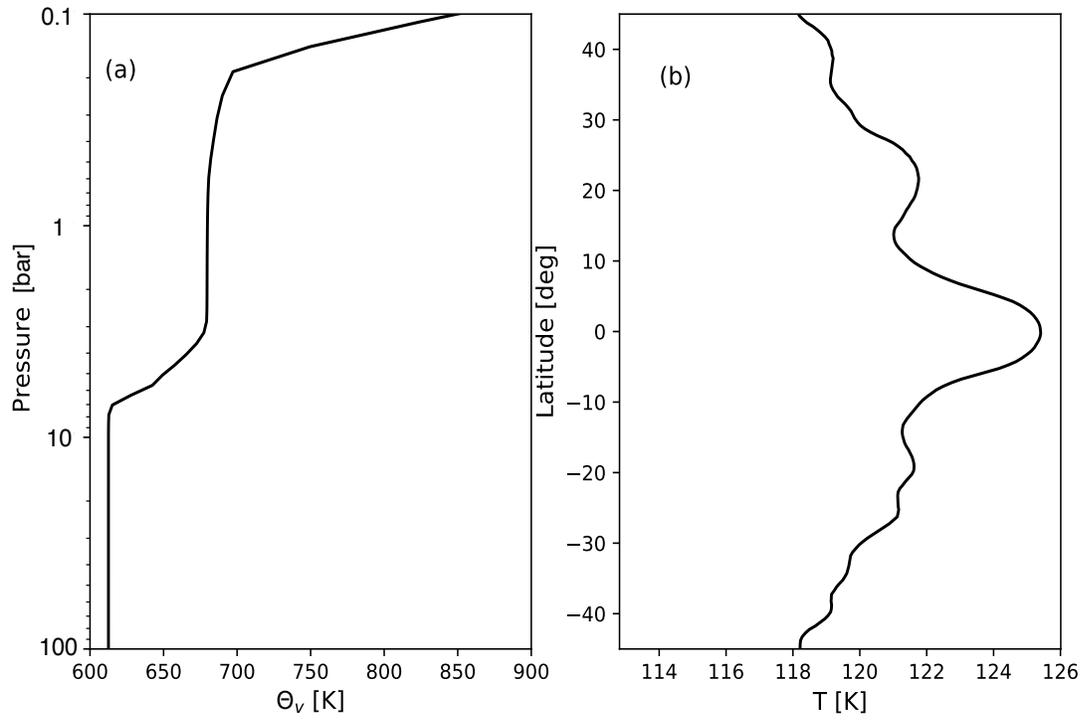

Figure 20. Global mean virtual potential temperature (a), zonal mean temperature at 0.25 bars (b).